\documentclass[a4paper,12pt]{article}

\pdfoutput=1 

\usepackage{jcappub} 

\def\ben{\begin{equation}}
\def\een{\end{equation}}

\def\bea{\begin{eqnarray}}
\def\eea{\end{eqnarray}}

\newcommand{\be}{\begin{eqnarray}}
\newcommand{\ee}{\end{eqnarray}}
\newcommand{\x}{x}
\newcommand{\xx}{\chi}
\newcommand{\m}{\mu}
\newcommand{\sign}{{\rm sign}}

\usepackage{xcolor}

\renewcommand{\theequation}{\arabic{section}.\arabic{equation}}

\begin{document}

\title{Weyl metrics and wormholes}   

\author[a,b]{Gary W. Gibbons,}
\emailAdd{gwg1@cam.ac.uk}
\affiliation[a]{DAMTP, University of Cambridge, Wilberforce Road, Cambridge CB3 0WA, UK}

\author[b,c]{Mikhail S. Volkov}
\emailAdd{volkov@lmpt.univ-tours.fr}

\affiliation[b]{Laboratoire de Math\'ematiques et Physique Th\'eorique,
LMPT CNRS -- UMR 7350, Universit\'e de Tours, 
Parc de Grandmont, 37200 Tours,
France
}
\affiliation[c]{%
Department of General Relativity and Gravitation, Institute of Physics,\\
Kazan Federal University, Kremlevskaya street 18, 420008 Kazan, Russia
}


\abstract{
We study solutions  obtained via applying 
dualities and complexifications 
to the vacuum Weyl metrics generated by massive rods and by point masses. 
 Rescaling them and extending to  complex  parameter values yields axially symmetric 
vacuum solutions containing    singularities along 
circles that can be viewed as singular matter sources.  
These solutions have  wormhole topology with several asymptotic regions interconnected by throats
and their sources can be viewed as thin rings of negative tension encircling the throats.
 For a particular value of the ring tension 
the geometry becomes exactly  flat although the topology remains non-trivial, so that the rings 
literally produce holes in flat space. To create a single ring wormhole of one metre radius one needs
a negative energy equivalent to the mass of Jupiter. 
Further duality transformations dress  the rings with the scalar field, 
either conventional or phantom. 
This gives rise to large classes of static, axially symmetric solutions,
presumably including all previously known solutions for a gravity-coupled massless scalar field,
as for example  the spherically symmetric  Bronnikov-Ellis wormholes with phantom scalar. 
The multi-wormholes contain infinite struts everywhere at the symmetry axes, apart from 
solutions with locally flat geometry. 
}


\maketitle

\section{Introduction}
\setcounter{equation}{0}

Wormholes are bridges or tunnels  between
different universes or different parts of the same universe. They were first introduced 
by Einstein and Rosen  (ER) \cite{Einstein:1935tc}, who noticed that the Schwarzschild black hole 
actually has two exterior  regions connected by a bridge\footnote{The analysis 
of Einstein and Rosen of 1935 can be related to the previous 
work of Flamm  \cite{Flamm}  of 1916  (see \cite{Flamm2015},\cite{FlammGibbons}
for the English translation) who was the first to construct the isometric embedding 
of the Schwarzschild solution. This relation is  discussed in the Appendix \ref{A0} below.}. 
The ER bridge is spacelike
and cannot be traversed by classical objects, but it has been argued   
 that it may connect quantum particles to produce quantum entanglement 
and the Einstein-Pololsky-Rosen (EPR) effect \cite{Einstein:1935rr}, hence ER=EPR \cite{Maldacena:2013xja}.
Wormholes were also considered as geometric models of elementary particles -- handles of space 
trapping inside an electric flux, say, which description  
may indeed be valid at the Planck scale  \cite{Misner:1957mt}. Wormholes can also describe initial data 
for the Einstein equations \cite {Misner:1960zz} (see \cite{Cvetic:2014vsa} for a recent review) 
whose time evolution 
corresponds to the black hole collisions of the type observed in the recent  
GW150914  event \cite{Abbott:2016blz}.

An interesting topic is traversable wormholes -- globally static bridges accessible for ordinary classical particles 
or light  \cite{Morris:1988tu} (see \cite{Visser:1995cc} for a review).  
In the simplest case such 
a wormhole is described by a static, spherically symmetric line element 
\be           \label{100}
ds^2=-Q^2(r)dt^2+dr^2+R^2(r)(d\vartheta^2+\sin^2\vartheta d\varphi^2),
\ee  
where $Q(r)$ and $R(r)$ are symmetric under $r\to -r$  and $R(r)$ 
attains a non-zero global 
minimum at $r=0$. If both $Q$ and $R/r$ approach unity as $r\to\pm\infty$ then 
the metric describes two asymptotically flat regions connected by a throat of 
radius $R(0)$.   The Einstein equations $G^\mu_\nu=T^\mu_\nu$
imply that the energy density $\rho=-T^0_0$ and the 
radial pressure $p=T^r_r$ satisfy  at $r=0$ 
\be         \label{200}
\rho+p=-2\frac{R^{\prime\prime}}{R}<0,~~~~~
p=-\frac{1}{R^2}<0.
\ee
It follows that for a static wormhole to be a solution of the Einstein equations,  
the Null Energy Condition (NEC), 
$T_{\mu\nu}v^\mu v^\nu=R_{\mu\nu} v^\mu v^\nu \geq 0$ for any null $v^\mu$, 
must be violated. 
Another  demonstration \cite{Morris:1988tu}  of the violation of the NEC uses the  
Raychaudhuri equation \cite{Hawking:1973uf} for a bundle of light rays described by
$\theta,\sigma,\omega$ : the expansion, shear and vorticity. 
In the spherically symmetric case one has  $\omega=\sigma=0$ \cite{Visser:1995cc}, 
hence 
\be        \label{4}
\frac{d\theta}{d\lambda}=-R_{\mu\nu}v^\mu v^\nu-\frac12\,\theta^2\,.
\ee
If rays pass through a wormhole throat, there is a moment of minimal cross-section area, $\theta=0$
but $d\theta/d\lambda>0$, hence $R_{\mu\nu}v^\mu v^\nu<0$ and the NEC is violated. 

If the spacetime is not spherically symmetric then the above arguments do not apply, 
but there are more subtle geometric considerations showing that the wormhole throat --
a compact two-surface of minimal area -- can exist if only the NEC is violated 
 \cite{Friedman:1993ty,Hochberg:1998ii}.
As a result,  traversable wormholes are possible if only the 
energy density becomes negative, for example due to vacuum polarization 
 \cite{Morris:1988tu}, or due to exotic matter 
as for example phantom fields with a negative kinetic energy 
\cite{Bronnikov:1973fh,Ellis:1973yv}.
Otherwise, one can 
search for wormholes in the alternative theories of gravity, as for example 
in the Gauss-Bonnet theory \cite{Kanti:2011jz},
in the brainworld  models 
\cite{Bronnikov:2002rn}, 
in  theories with 
non-minimally coupled fields 
\cite{Sushkov:2011jh},
or in massive (bi)gravity \cite{Sushkov:2015fma}. 

The best known wormhole solutions were found in 1973 in the theory of gravity-coupled 
phantom scalar field $\psi$ \cite{Bronnikov:1973fh,Ellis:1973yv}. 
Their geometry is {\it ultrastatic}, 
$ds^2=-dt^2+dl^2$, where the 3-metric $dl^2=\gamma_{ik}dx^i dx^k$ satisfies 
equations
\bea                                \label{Rp}
\overset{(3)}{R}_{ik}(\gamma)=-2\partial_i\psi\partial_k\psi,~~~~~~~~\Delta\psi=0,
\eea
the 
phantom nature of $\psi$ being  encoded in the negative sign in the right hand side of the 
gravitational  equations. 
The simplest solution of these equations is the 
spherically symmetric 
Bronnikov-Ellis (BE) wormhole \cite{Bronnikov:1973fh,Ellis:1973yv}, 
\be                       \label{BEw}
dl^2=dx^2+(x^2+\mu^2)(d\vartheta^2+\sin^2\vartheta d\varphi)^2,~~~~~~\psi=\arctan\left(\frac{x}{\mu}  \right),
\ee
where $\mu$ is an integration constant determining the size of the wormhole throat. 
More general solutions of \eqref{Rp} were obtained by Cl\'ement  \cite{Clement:1983tu} 
choosing the 3-metric to be in the 
axially symmetric Weyl form \cite{Weyl},
\be               \label{Weyl0}
dl^2=e^{2k}(d\rho^2+dz^2)+\rho^2 d\varphi^2,
\ee
where $k$ and $\psi$ depend on $\rho,z$. Equations \eqref{Rp} then reduce to
%
\bea              \label{eqWeyl0}
{\cal D}k=
-\rho\left({\cal D} \psi  \right)^2,~~~~~~~~
\frac{1}{\rho}\frac{\partial}{\partial\rho}\left(\rho\,\frac{\partial\psi}{\partial\rho}\right)+
\frac{\partial^2 \psi}{\partial z^2}=0,
\eea
where ${\cal D}=\partial_\rho+i\partial_z$. 
Since the $\psi$-equation is linear, 
superposing 
its elementary one-wormhole solutions gives solutions describing 
multi-wormholes \cite{Clement:1983tu,Clement:2015lul}
(they can be generalized to include also rotation and Maxwell field 
\cite{Clement:1983ic,Clement:1983ib}). 

It is worth noting that equations \eqref{Rp}
actually coincide with the {vacuum} Einstein equations 
for the everywhere non-singular 5-metric \cite{Clement:1983fe}
\be                         \label{5D}
ds_5^2=\cos(2\psi)[-dx_0^2+dx_4^2]+2\sin(2\psi)dx_0 dx_4+dl^2\,.
\ee
Therefore, wormhole solutions of \eqref{Rp} 
can be interpreted as 5-geometries
without invoking the negative energy phantom\footnote{The BE solution \eqref{BEw},
when lifted to 5D, becomes a member of spherically symmetric vacuum solutions of
Chodos and Deweiler \cite{Chodos:1980df}.}.

Wormholes are usually  associated with exotic matter and 
it is  little known that they exist also in General Relativity (GR)
as classical solutions of {\it vacuum} Einstein equations. For such solutions the metric 
is Ricci flat everywhere apart from geometric sets of zero measure that can be 
interpreted as singular ring-shaped matter sources whose effective energy-momentum tensor
has the structure needed to provide the NEC violation.  Since these solutions 
are vacuum everywhere apart from singularities,  they can be constructed 
with the standard GR methods. The first example  of this type was discovered in 1966 by 
Zipoy who studied {\it oblate} vacuum metrics of the axially symmetric Weyl type \cite{Weyl}
and found solutions which ``have a ring singularity and have a double sheeted topology; 
one can get from one sheet to the other by going through the ring"   \cite{Zipoy}. The two sheets 
are the two asymptotically flat regions connected by the wormhole throat -- the hole encircled by the ring.
As Zipoy noted, this hole should have ``strange properties:  an object falling through it would not be seen 
coming out from the other side, whereas if viewed from above the ring it could be seen dropping through"  \cite{Zipoy}.
The ring singularity can be interpreted in this case 
as a cosmic string loop with the {\it negative} tension 
\cite{Bronnikov:1997gj,Clement:1998nk,Clement:1998yt,Gibbons:2016bok}, 
\be
T=-\frac{(1+\sigma^2)c^4}{4G},
\ee 
where  $\sigma$ is a free parameter. Interestingly, if $\sigma=0$ then the metric  becomes {\it flat} everywhere 
outside the ring, the latter thus literally creating a hole in flat space \cite{Gibbons:2016bok}. 

Vacuum ring wormholes may seem to be very different from the BE wormholes supported by the phantom field. 
However, they are actually described by the same solution. Specifically, the ring metric is 
\be
ds^2=-e^{2U}dt^2+e^{-2U}dl^2
\ee
with $dl^2$ given by \eqref{Weyl0}, and the vacuum Einstein equations read in this case
\bea              \label{eqWeyl1}
{\cal D}k=
+\rho\left({\cal D} U  \right)^2,~~~~~~~~
\frac{1}{\rho}\frac{\partial}{\partial\rho}\left(\rho\,\frac{\partial U}{\partial\rho}\right)+
\frac{\partial^2  U}{\partial z^2}=0.
\eea
Now, comparing with \eqref{eqWeyl0} it is clear that solutions of the two sets of equations 
are related by 
\be                  \label{Uk}
U\leftrightarrow\psi,~~~~~~~k\leftrightarrow -k.
\ee
Therefore, ring wormholes and phantom wormholes are actually described by the same equations 
and to obtain ones from the others it is enough to 
interchange the phantom $\psi$ with the Newtonian potential $U$ and to flip the sign of $k$. 
As a result, to study wormholes one can  use the ordinary vacuum GR without invoking 
phantom fields or extra dimensions. The price to pay is the metric singularity
supported by the ring which provides the necessary NEC violation.

In this paper we study solutions  obtained via applying 
dualities and complexifications 
to the vacuum Weyl metrics.
 As a starting point we use the elementary metrics  generated by rods and by point masses. 
 Rescaling them and extending to the  complex 
 parameter values yields the {\it prolate} and {\it oblate} vacuum solutions. 
Further duality transformations produce a scalar field, 
which can be either conventional or phantom. 
This gives rise to large classes of solutions,
presumably including all previously known static, axially symmetric 
solutions for a gravity-coupled massless scalar field\footnote{We do not consider stationary or 
time-dependent solutions; see, e.g., \cite{Clement:1983ib,Tahamtan:2016fur}.}.
Especially interesting are the oblate solutions which, irrespectively of whether 
they support a scalar or not, describe wormholes connecting several 
asymptotic regions.  In the one-wormhole sector they reduce to the ring wormholes 
in the vacuum case, and to the Bronnikov-Ellis wormhole in the phantom case.

The main point we would like to emphasise  is that wormholes exist already in vacuum 
GR.  This was known already to Zipoy but remains largely unknown to the community  
even now.  The vacuum wormholes may be viewed as ``primary'' since those 
sourced by a scalar can be obtained from them by duality rotations. 
We shall therefore summarize the essential facts and give 
a description of the vacuum wormholes, 
before applying  the duality rotations to produce more general solutions with scalar.

The rest of the text is organised as follows. In Section II we introduce the theory of gravitating 
massless scalar and describe its simplest solutions, both in the case of  conventional scalar $\phi$ 
and for the phantom  $\psi$. In Section III we impose the axial symmetry and list the duality transformations
which map solutions to solutions. Section IV describes the simplest vacuum Weyl metrics
generated by massive rods and by point masses. The vacuum wormholes 
obtained by rescaling and complexifying the one-rod solution are discussed in Section V,
where their topology and the structure of the ring source are considered. 
 This Section  also describes  solutions 
with scalar obtained via applying the dualities.
Milti-wormholes are considered in Section VI, where we also analyse  carefully the regularity at the symmetry axes 
and describe the two ring and the $N$-ring solutions with locally 
flat geometry. Finally, we consider in Section VII solutions obtained from the 
Chazy-Curzon metrics and conclude in Section VIII. 
The isometric embeddings for the BE wormhole are described in  Appendix~\ref{A},
while the relation between work of Famm and that of Einstein-Rosen is discussed in Appendix~\ref{A0}. 

Unless otherwise stated, Planck units are used 
everywhere. A short version of this text can be found in \cite{Gibbons:2016bok}.

\section{Gravitating scalar field}
\setcounter{equation}{0}

We consider a system with a real gravitating scalar field, 
\be
{\cal L}=[R-2\epsilon\, (\partial \Phi)^2]\sqrt{-g}\,.
\ee
Here the parameter  takes two values, either 
$\epsilon=+1$ corresponding to the conventional scalar field,
or $\epsilon=-1$ corresponding to the phantom field. If $\epsilon=+1$ then 
the energy is positive and the scalar field mediates
an attractive force. If $\epsilon=-1$ then the scalar kinetic energy is negative
(although the total energy may still be positive) and the corresponding force is
repulsive.  In what follows we shall be mainly interested in the phantom case,
however, it is instructive to consider the two cases together. 

Assuming the spacetime metric to be static 
\be                    \label{static}
ds^2=-e^{2U} dt^2+e^{-2U}\gamma_{ik}dx^i dx^k
\ee
where $U$, $\gamma_{ik}$ and $\Phi$ are functions of the spatial coordinates $x^k$, the Lagrangian 
becomes
\be
{\cal L}=\left[\stackrel{(3)}{R}-2 \gamma^{ik}(\partial_i U\partial_k U+\epsilon\,\partial_i \Phi\partial_k \Phi)\right]\sqrt{\gamma}\,.
\ee
The field equations are 
\bea                    \label{eq}
\frac12 \stackrel{(3)}{R}_{ik}&=&\partial_i U \partial_k U +\epsilon\, \partial_i\Phi\partial_k \Phi \,,\nonumber \\
\Delta U&=&0\,,\nonumber \\
\Delta \Phi &=0\,.
\eea
In what follows we shall 
 denote $\Phi=\phi$ if $\epsilon =1$ and $\Phi=\psi$ if $\epsilon=-1$. Formally, 
 the change $\epsilon=+1\to \epsilon=-1$ can  be achieved 
 by extending  the scalar field  to purely imaginary values: $\phi\to i\psi$. 

Equations \eqref{eq} admit the following symmetry. 
If $\epsilon=1$ and $(U,\phi,\gamma_{ik})$  is a solution then 
the following replacement will give solutions:
\bea                    \label{1}
U&\to& U\,\cos\alpha + \phi\,\sin\alpha\,, \nonumber \\
\phi&\to& \phi\,\cos\alpha-U\,\sin\alpha\,,\nonumber \\
\gamma_{ik}&\to&\gamma_{ik}\,,
\eea
where $\alpha$ is a constant parameter. 
Likewise, if $\epsilon=-1$ and $(U,\psi,\gamma_{ik})$  is a solution then 
other solutions can be obtained via 
\bea                   \label{2}
{U}&\to& U\,\cosh\alpha + \psi\,\sinh\alpha\,, \nonumber \\
{\psi}&\to& \psi\,\cosh\alpha+U\,\sinh\alpha\,, \nonumber \\
\gamma_{ik}&\to& \gamma_{ik}. 
\eea
Many non-trivial solutions  can be generated by applying these symmetries,
let us consider simple  examples.

\subsection{Spherically symmetric sector}

All known static, spherically symmetric solutions with scalar can be obtained 
from the vacuum Schwarzschild solution  written in the form 
(here $d\Omega^2=d\vartheta^2+\sin^2\vartheta d\varphi^2$)
\bea
ds^2&=&-\frac{\x-m}{\x+m}\,dt^2+\frac{\x+m}{\x-m}\,d\x^2+(\x+m)^2d\Omega^2 \nonumber 
\eea
by applying the rotations  \eqref{1} or boosts \eqref{2}. 
Comparing with the line element \eqref{static} yields
\be                                  \label{S}
e^{2U}=\frac{\x-m}{\x+m},~~~~~~~\gamma_{ik}dx^i dx^k=d\x ^2+(\x^2-m^2)d\Omega^2,~~~~~\Phi=0. 
\ee

\subsubsection{Solutions with  scalar field $\phi$}
Let us choose $\epsilon=+1$, $\Phi=\phi$. 
Applying to \eqref{S} 
the rotation \eqref{1} with $\cos\alpha=1/s$ 
gives 
\be
e^{2U}=\frac{\x-m}{\x+m}\to \left(\frac{\x-m}{\x+m} \right)^{{1}/{s}},~~~~~~~~
e^{2\phi}=1\to \left(\frac{\x -m}{\x +m} \right)^{{\sqrt{s^2-1}}/{s}},
\ee
while the 3-metric $\gamma_{ik}$ does not change. Therefore, 
\bea                               \label{F}
ds^2&=&-\left(\frac{\x-m}{\x+m} \right)^{{1}/{s}}dt^2+
\left(\frac{\x+m}{x-m} \right)^{{1}/{s}}\left[
dx^2+(\x^2-m^2)d\Omega^2
\right], \nonumber \\
\phi&=&\pm \frac{\sqrt{s^2-1}}{2s}\ln\left(\frac{\x -m}{\x +m} \right),~~~~~~~~~|s|\geq 1,
\eea
which are the well known solutions found by Fisher \cite{Fisher:1948yn} and by 
Janis-Robinson-Winicour 
\cite{Janis:1968zz} (FJRW)\footnote{The $d$-dimensional version of this solution was discussed 
in \cite{Abdolrahimi:2009dc}.}.
If $s\to\infty$ then the metric becomes ultra-static, 
\bea                   \label{swap0}
ds^2=-dt^2+dx^2+(x^2-m^2)\,d\Omega^2,~~~~~~~\phi=\pm\frac12\ln\left(\frac{\x -m}{\x +m} \right).
\eea

\subsubsection{Solutions with phantom  $\psi$}
Let us now set $\epsilon=-1$, $\Phi=\psi$ and apply 
to \eqref{S}  the boost \eqref{2}
with $\cosh\alpha=1/s$. This gives the ``phantom version" of the FJRW solutions \eqref{F},  
\bea                               \label{FF}
ds^2&=&-\left(\frac{\x-m}{\x+m} \right)^{{1}/{s}}dt^2+
\left(\frac{\x+m}{\x-m} \right)^{{1}/{s}}\left[
dx^2+(\x^2-m^2)d\Omega^2
\right], \nonumber \\
\psi&=&\pm \frac{\sqrt{1-s^2}}{2s}\ln\left(\frac{\x-m}{\x+m} \right)~~~~~~~~~~~~~|s|\leq 1. 
\eea
Although 
the metric looks  identical to that in  \eqref{F}, 
the geometry is different because 
the range of $s$ is now different.
One may think that these solutions could describe regular black holes, since setting $1/s$ to integer values 
removes the branching singularity at the horizon $x=m$. However, the area of the sphere
is proportional to 
\be
(x-m)^{1-1/s}\,,
\ee
hence the horizon has an infinite area, therefore these solutions cannot describe regular black holes.

\subsubsection{Wormholes}

Let us extend the parameters $m,s$ in \eqref{F}, \eqref{FF} to imaginary values, 
\be                     \label{comp}
m\to i \mu,~~~~~s\to- i s.
\ee
One has then 
\be
\frac{\x-m}{\x+m}\to \frac{\x-i\m}{\x+i\m}=e^{-2i\Psi}\,,~~~~~~\Psi=\arctan\left(
\frac{\x}{\m}\right),
\ee
hence 
the metric in  \eqref{F}, \eqref{FF} remains real, 
the scalar field in \eqref{F} becomes
purely imaginary, $\phi\to i\psi$, while the phantom field in \eqref{FF} remains real, $\psi\to\psi$. 
As a result, upon the complexification \eqref{comp} 
both the FJRW solutions \eqref{F} and their phantom counterparts \eqref{FF} reduce 
to the same solutions for the phantom field, 
\bea                                  \label{W}
ds^2&=&- e^{2\Psi/ s}dt^2 +e^{-2\Psi/ s}[dx^2+(x^2+\m^2)d\Omega^2]\,, \nonumber \\
\psi&=&\pm \frac{\sqrt{s^2+1}}{s}\,\Psi.
\eea
These solutions describe 
globally regular and asymptotically flat wormholes  
(see the Appendix \ref{A} where the isometric imbedding of the geometry \eqref{W} into 
a higher-dimensional Minkowski space is considered). 
They owe their existence to the repulsive nature of the phantom field --
the gravitational attraction being compensated by the scalar field repulsion gives rise to 
globally regular equilibrium configurations. 
In the $ s\to\infty$ limit one obtains the ultra-static wormhole of Bronnikov and Ellis 
\cite{Bronnikov:1973fh,Ellis:1973yv},
\bea                                \label{UW}
ds^2&=&- dt^2 +dx^2+(x^2+\m^2)\,d\Omega^2\,, ~~~~~~\psi=\pm \Psi.
\eea
This  can also be obtained directly from the ultra-static solution \eqref{swap0} 
via $m\to i\m$.

\subsection{Anti-gravitating   solutions with phantom field  \label{antigr}}

Other  solutions for the phantom field can be constructed  in view of the following 
observation \cite{Gibbons:2003yj}.  Taking the infinite boost limit in \eqref{FF},
\be
s\to 0,~~m\to 0,~~~~~M=-\frac{m}{s}=const,
\ee
gives 
\bea
ds^2&=&-e^{2U}+e^{-2U}(dx^2+x^2 d\Omega^2), ~~~~~~
U=\frac{M}{x}\,,~~~~\psi=\pm U.
\eea
One notices that the 3-metric $\gamma_{ik}$ becomes flat and 
$\psi=\pm U$ is a harmonic function. This allows one to generalize the 
solutions because if  $U=\pm \psi$ then Eqs.\eqref{eq} show that 
$\stackrel{(3)}{R}_{ik}=0$ and hence one can choose $\gamma_{ik}=\delta_{ik}$. 
The equation for $U$ then becomes $\Delta U=0$ where $\Delta$ is the ordinary 
Laplace operator. Therefore, any harmonic function $U=\pm \psi$ gives a solution, 
for example  the multi-centre solution,  
\bea
ds^2&=&-e^{2U}+e^{-2U} (d {\bf x})^2 ,~~~~~~
U=\sum_{n=1}^N \,\frac{M_n}{|{\bf x}-{\bf x}_n|},~~~~~~~~\psi=\pm U, 
\eea
where $M_n$ and ${\bf x}_n$ are arbitrary. 
The existence of such solutions is the speciality  of the phantom field
whose repulsion can exactly compensate the gravitational attraction.

\section{Axially symmetric Weyl metrics and their symmetries}
\setcounter{equation}{0}

Let us return to equations \eqref{eq} and assume the system to be axially symmetric.
Then the spacetime metric can be put to the Weyl form, 
\be                   \label{www}
ds ^2 = - e^{2U}dt ^2 + e^{-2U} \bigl \{  e^{2k} \bigl( d \rho ^2 + dz ^2 \bigr ) + \rho ^2 d \varphi ^2 \bigr \}, 
\ee
where $U,k$ and $\Phi$ depend on $\rho,z$. 
Then equations \eqref{eq} reduce to\footnote{One can check that this reduction is indeed consistent, which
would not be the case if the scalar field had a potential. The Weyl 
formulation is also consistent for an electrostatic vector field, so that it applies, for example,
within the electrostatic sector of  dilaton gravity.}
\be                  \label{drop}
\frac{\partial^2 U}{\partial\rho^2}&+&\frac{1}{\rho}\,\frac{\partial U}{\partial\rho}+\frac{\partial^2 U}{\partial z^2}=0, \nonumber \\
\frac{\partial^2 \Phi}{\partial\rho^2}&+&\frac{1}{\rho}\,\frac{\partial \Phi}{\partial\rho}+\frac{\partial^2 \Phi}{\partial z^2}=0, \nonumber \\
\frac{\partial k}{\partial\rho}&=&\rho\left[
\left(\frac{\partial U}{\partial\rho}\right)^2
-\left(\frac{\partial U}{\partial z}\right)^2
+\epsilon
\left(\frac{\partial \Phi}{\partial\rho}\right)^2
-\epsilon\left(\frac{\partial \Phi}{\partial z}\right)^2
\right], \nonumber \\
\frac{\partial k}{\partial z}&=&2\rho\,\left[
\frac{\partial U}{\partial\rho}
\frac{\partial U}{\partial z} +\epsilon\,
\frac{\partial \Phi}{\partial\rho}
\frac{\partial \Phi}{\partial z}
\right]. 
\ee 
Here the last two equations are compatible with each other in view of the first two equations; 
their solution can be obtained from the line integral 
\be                  \label{kkkk}
k=\int\left(
\frac{\partial k}{\partial\rho}\,d\rho+\frac{\partial k}{\partial z}\,dz
\right).
\ee
This expresses, in principle, $k$ in terms of $U,\Phi$, although  in practice the computation of the integral
may be difficult. 
Since $\partial_zk\sim\rho$ vanishes at $\rho=0$
(if only $U,\Phi$ are non-singular there), it follows that 
$k=k_0$ at the symmetry axis.  
If $k_0=0$ then the geometry 
at the axis will be regular, but  
for $k_0\neq 0$   there is a conical singularity
on that portion of the axis. Physically this corresponds to a cosmic strut or a
cosmic string with deficit angle 
\ben               
\delta = 2 \pi (1-e^{-k_0})   
\een 
giving rise to a force or tension
\ben
T=  \frac{\delta}{8 \pi}. 
\een
Such conical singularities are typical for Weyl solutions. 

\subsection{Symmetries} 

Equations \eqref{drop} are invariant under the following operations. 

\subsubsection{Sign flips and scaling}
If $(U,k,\Phi)$ is a solution of Eqs.\eqref{drop}
then the following replacements will give solutions: 
\bea
U\to U,~~~~&k\to k&,~~~~~~ \Phi\to -\Phi;  \nonumber  \\
U\to -U,~~~~&k\to k&,~~~~ \Phi\to \Phi.
\eea
These symmetries are not very interesting:
the first one simply changes sign of the scalar field, while the second one,
when applied to solutions \eqref{F},\eqref{FF},\eqref{W}, 
changes sign of the parameters $m$ and $\m$. 
A more interesting symmetry is 
\bea                       \label{scale}
U\to \lambda U,~~~~k\to \lambda^2 k,~~~~ \Phi\to \lambda\Phi,  
\eea
with constant $\lambda$. 
As we shall see, this symmetry acts in a non-trivial way completely changing properties 
of solutions. 

These three symmetries do not mix up gravitational and scalar variables 
so that, in particular, vacuum solutions with $\Phi=0$ remain vacuum. 
There are also symmetries which intermix the variables, they are different 
for the ordinary scalar field $(\epsilon=+1)$ and for the phantom field $(\epsilon=-1)$.  

\subsubsection{Rotations and boosts}
These symmetries generate solutions with scalar field.  
The rotations  apply only in the conventional scalar case
($\epsilon=+1$  and $\Phi=\phi$) acting on a solution
 $(U,k,\phi)$ as
\bea                                   \label{sym1}
U\to U \cos \alpha + \phi \sin \alpha,~~~~k\to  k,~~~~ \phi\to \phi \cos \alpha -U \sin \alpha,
\eea
where $\alpha$ is a constant parameter. The boosts 
 apply only in the phantom case
($\epsilon=-1$  and $\Phi=\psi$)
acting on 
 $(U,k,\psi)$ as
\bea                                   \label{sym2}
U\to U \cosh \alpha + \psi \sinh \alpha,~~~~k\to  k,~~~~ \psi\to \psi \cosh \alpha +U \sinh \alpha. 
\eea

\subsubsection{Swap symmetry}
This applies 
only in the phantom field case, 
\bea                                   \label{swap}
U\to \psi,~~~~k\to  -k,~~~~ \psi\to U,
\eea
so that $U$ and $\psi$ swap while $k$ flips sign. 
The existence of this symmetry implies that there are twice as many solutions with phantom field 
as those with the ordinary scalar. In particular, this symmetry relates the ring 
wormholes and the phantom wormholes. 

\subsubsection{Tachyon symmetry} 
A complex change of the temporal and azimuthal variables 
\be
t\to i\varphi,~~~~~~\varphi\to i t 
\ee
puts  the spacetime metric \eqref{www} to the form 
\be                   \label{www1}
ds ^2 = - \rho^2 e^{-2U}dt ^2 + e^{-2U} \bigl \{  e^{2k} \bigl( d \rho ^2 + dz ^2 \bigr )\bigr \}  + e^{2U} d \varphi ^2,
\ee
which is again the Weyl metric obtained from the original one via
\bea                                   \label{tachyon}
U\to \ln\rho-U,~~~~k\to  k-2U+\ln\rho,~~~~ \Phi\to \Phi.
\eea
One can check that this is a symmetry of equations \eqref{drop} which does not change the scalar. 
This symmetry has been relatively little studied. 
Applied to the Schwarzschild metric it produces 
the Peres-Schulman-Gott
(PSG) tachyon
 solution with
isometry group $SO(2) \times SO(2,1)$  
\cite{Peres}, \cite{Schulman:1971ap}, \cite{Gott:1974yc}. 
This is  appropriate for a neutral tachyon moving in a spacetime
 with one spatial direction  
compactified; it describes the aftermath
of  vacuum decay \cite{Gibbons:1996pd}.

\section{Vacuum Weyl metrics}

Let us describe the simplest solutions of Eqs.\eqref{drop} with $\Phi=0$. 

\subsection{One-rod solution}

The Schwarzschild solution 
$$
ds^2=-e^{2U}dt^2+e^{-2U} dl^2
$$
with
\be           
e^{2U}=\frac{\x-m}{\x+m},~~~~~~~~dl^2\equiv \gamma_{ik}dx^i dx^k=d\x^2+(\x^2-m^2) d\Omega^2\,
\ee
can be 
transformed to the Weyl form. The transformation is achieved by setting 
\be                                          \label{x}
z=x\cos\vartheta,~~~~~~\rho=\sqrt{x^2-m^2}\sin\vartheta,
\ee
hence 
\be                            \label{rz}
d\rho^2+dz^2=\frac{x^2-m^2\cos^2\vartheta}{x^2-m^2}\left[dx^2+(x^2-m^2)d\vartheta^2\right], 
\ee
which gives 
\be
ds^2=-e^{2U} dt^2+e^{-2U}\left[e^{2k}(d\rho^2+dz^2)+\rho^2 d\varphi^2\right],
\ee
with
\be                           \label{BOT}
e^{2U}=\frac{\x-m}{\x+m},~~~~~~~~~
e^{2k}=\frac{\x^2-m^2}{\x^2-m^2\cos^2\vartheta}. 
\ee
There remains to express $U,k$ in terms of $\rho,z$. 
Inverting  \eqref{x} gives $x,\vartheta$ in terms of $z,\rho$, 
\be
x=R,~~~~~~x\pm m\cos\vartheta=R_\pm,
\ee
where
\be               \label{SWW}
R =\frac12(R_{+}+R_{-}), ~~~~ ~~~~~R_{\pm}=\sqrt{\rho^2+(z\pm m)^2}.
\ee
Injecting this to \eqref{BOT} gives $U,k$ expressed in terms of Weyl coordinates, 
\be                                       \label{SW}
U=\frac12\,\ln\left(\frac{R-m}{R+m}\right),~~~~~~~~~~~
k=\frac12\ln\left(
\frac{R^2-m^2}{R_{+}R_{-}}
\right),
\ee
and one can directly check that these fulfill 
equations \eqref{drop}. 
This gives the Schwarzschild metric in the Weyl form. 
One notices  that $U$ can be represented as 
the Newtonian potential of a thin rod of 
linear mass density $1/2$ and of mass $m$, hence\footnote{Directly calculating the integral in \eqref{U} 
gives
$U=\frac12\ln\frac{R_{+}-z-m}{R_{-}-z+m}$ which is equivalent to $U$ in \eqref{SW}. }
\be
U                \label{U}
=-\frac12\int_{-m}^{m}\frac{d\zeta}{\sqrt{\rho^2+(z-\zeta)^2}}\,.
\ee
For this solution  one has $k(\rho=0,z)=0$ if $|z|>m$
therefore  there are no struts and the geometry is regular on parts of the axis not occupied  by the rod.

\subsection{Two-rod solution}
More general solutions can be obtained by taking several rods and superposing 
their Newtonian potentials. This gives $U$, while $k$ is  obtained 
from  \eqref{kkkk}. 
In the simplest two-rod case one has \cite{IsraelKhan}
\be                                  \label{2rod}
U=U_1+U_2,~~~~~~~~k=k_1+k_2+k_{12}\,,
\ee
where (with $a=1,2$)
\be            \label{TR}
U_a&=&\frac{1}{2}\ln\left(
\frac{R_a-m_a}{R_a+m_a}
\right),~~~~~~~~k_a=\frac12\ln\left( \frac{(R_a)^2-(m_a)^2}{R_{a+}R_{a-}  } \right),~~~\nonumber \\
k_{12}&=&\frac12\ln\left(
\frac{ (R_{1+}R_{2-}+z_{1+}z_{2-}+\rho^2) (R_{1-}R_{2+}+z_{1-}z_{2+}+\rho^2) }{
(R_{1+}R_{2+}+z_{1+}z_{2+}+\rho^2)(R_{1-}R_{2-}+z_{1-}z_{2-}+\rho^2)   }
\right),
\ee
with
\be
z_{a\pm}=z-z_a\pm m_a,~~~~~~~~R_{a\pm}=\sqrt{\rho^2+(z_{a\pm})^2}\,,~~~~~R_a=\frac12(R_{a+}+R_{a-}).
\ee
For this solution $k$ vanishes on parts of the $z$-axis non-occupied by the rods,
except for the interval between the rods where it is constant and negative. 
Therefore, there is a negative tension
strut between the two rods giving rise to the repulsive force. This strut ``props up" the two black holes
and does not let them fall on each other. 

The generalization to $N$ rods is straightforward \cite{IsraelKhan},
\be                                  \label{Nrod}
U=\sum_{a=1}^N U_a,~~~~~~~~k=\sum_{a=1}^N k_a+\sum_{a<b}k_{ab}\,,
\ee
where $U_a,k_a$ are given by the same expressions as in \eqref{TR}, while $k_{ab}$ is obtained from 
$k_{12}$ in  \eqref{TR} by replacing $R_{1\pm}\to R_{a\pm}$ and $R_{2\pm}\to R_{b\pm}$.

\subsection{The Chazy-Curzon metrics}

Instead of the Newtonian potential of rods one can consider that 
of point masses located at the $\rho=0$ axis. For just one mass $m$ located at $z=0$ 
one obtains the solution
\be                        \label{Curz}
U= - \frac{m}{R}\,, \qquad
 k =-\frac{m^2\rho ^2}{2 R^4 } \,,
\ee
where $R=\sqrt{\rho^2+z^2}$. 

For two masses $m_\pm$  located $z=\pm m$ one has, with 
$R_\pm=\sqrt{\rho^2+(z\pm m)^2}$,
\bea                \label{Curz1}
U&=& - \frac{m_{+}}{R_{+}}- 
\frac{m_{-}}{R_{-}},  \nonumber \\ 
 k &=&- \frac{m_{+}^2 \rho ^2}{2 (R_{+})^4 }
- \frac{m_{-}^2 \rho ^2}{2 (R_{-})^4 }+
 \frac{m_{+}m_{-}}{2m^2} \left(
 \frac{ \rho^2 + z^2-m^2}  
{  R_{+} R_{-} } -1\right). 
\eea
These solutions were constructed by Chazy \cite{Chazy1924} and Curzon \cite{Curzon1924}
and later independently by  Silberstein \cite{Silberstein:1936zz}. The one-mass solution
is regular at the axis, while for the two masses one has  
\ben
k(\rho=0,z)=\frac{m_{+}m_{-}}{2m^2} \left(
 \frac{z^2-m^2}  
{|z^2-m^2|} -1\right). 
\een
This vanishes outside the interval between the two particles where 
$z^2-m^2 >0$ but not inside this interval where
$z^2-m^2 <0$ and hence \cite{Schleifer} 
\ben
k(\rho=0,z)=-\frac{m_{+}m_{-}}{m^2}.
\een 
The original paper of Curzon \cite{Curzon1924}\footnote{The  reader is warned 
that reference to Curzon's paper  \cite{Curzon1924} 
in the literature is often given incorrectly.
Interestingly, this paper   of 1924 also gives what is now known 
 as the Majumdar-Papapetrou
\cite{Papapetrou,Majumdar:1947eu} 
multi-centre solution of Einstein -Maxwell theory.}
also gives the solution for $N$ point masses.%

\section{Solutions  from one rod}

\setcounter{equation}{0}

We described above the elementary vacuum solutions generated by rods and by point masses. 
One could now apply rotations  \eqref{sym1} and boosts \eqref{sym2} to produce 
new solutions with scalar field. 
However, before doing this we notice that yet much larger families of 
vacuum metrics can be obtained via acting first with the scale symmetry \eqref{scale} 
and then performing complexification. This gives the {\it prolate} and {\it oblate} vacuum metrics.

\subsection{Prolate vacuum metrics}

Applying the scaling \eqref{scale} to the one-rod Schwarzschild solution \eqref{SW} gives 
the two-parameter family of solutions labeled by $\lambda$ and $m$, 
\be                                       \label{SW1}
U=\frac{\lambda}{2}\,\ln\left(\frac{R-m}{R+m}\right),~~~~~~~~~~~
k=\frac{\lambda^2}{2}\ln\left(
\frac{R^2-m^2}{R_{+}R_{-}}
\right),
\ee
with 
\be                            \label{RRR}
R =\frac12(R_{+}+R_{-}),~~~~R_{\pm}=\sqrt{\rho^2+(z\pm m)^2}.
\ee
Using \eqref{x}, \eqref{rz}, \eqref{BOT} to pass back to the $x,\vartheta$ coordinates  
gives 
\bea                        \label{ZV}
ds^2&=&-\left(
\frac{x-m}{x+m}
\right)^\lambda dt^2+\left(
\frac{x-m}{x+m}
\right)^{-\lambda} dl^2,   \\
dl^2&=&\left(
\frac{\x^2-m^2\cos^2\vartheta  }{x^2-m^2  }
\right)^{1-\lambda^2}
\left[d\x^2+(\x^2-m^2) d\vartheta^2\right]+(x^2-m^2)\sin^2\vartheta d\varphi^2\,.  \nonumber 
\eea
These are  the {\it prolate} Zipoy-Voorhees (ZV) solutions \cite{Zipoy,Voorhees:1971wh}. 
Unless for $\lambda^2=1$ they are not spherically symmetric and exhibit a curvature 
singularity at $R=x=m$. 
These metrics have been relatively well studied, as they 
can be used to describe deformations of the Schwarzschild black hole 
(see for example \cite{Kodama:2003ch,LukesGerakopoulos:2012pq}). 
For all these solutions the relation between the $\x,\vartheta$ and $\rho,z$ coordinates is given by \eqref{x}, 
hence 
\be                        \label{x1}
 \frac{z^2}{\x^2}+\frac{\rho^2}{\x^2-m^2}=1.
\ee
Therefore, lines of constant $\x$ are {\it prolate} ellipses with the major semi-axis oriented along the $z$-direction. 
In the $x\to m$ limit these ellipses shrink to the $-m\leq z\leq m$ segment of the $z$-axis. 
Both $(\rho,z)$ and $(\x,\vartheta)$ coordinate cover the $R=x>m$ region of the manifold.

\subsection{Oblate vacuum metrics} 

Let us  continue the solution \eqref{SW1} to imaginary parameter values, 
\be                      \label{an}
m\to i\mu\,,~~~~~~~~~~~\lambda \to i\sigma.
\ee
In this case instead of \eqref{x} one will have 
\be                        \label{x2}
z=\x\cos\vartheta,~~~~~~
\rho=\sqrt{\x^2+\m^2}\,\sin\vartheta,
\ee
hence
\be
 \frac{z^2}{\x^2}+\frac{\rho^2}{\x^2+\m^2}=1. 
\ee
Therefore, lines of constant $\x$ in the $(\rho,z)$ plane 
are  {\it oblate} ellipses with the major semi-axis oriented along the $\rho$-direction. 
In the $x\to 0$ limit these ellipses shrink to the segment of the $\rho$-axis
\be                \label{opt1a}
{\cal I}=\{\rho\in[0,\mu],z=0\}
\ee
(assuming that $\mu>0$). 
The functions $R_\pm$ in \eqref{RRR} become complex-valued, 
\be                    \label{Xpm}
R_{\pm}=\sqrt{\rho^2+(z\pm m)^2}~~\to ~~
\sqrt{\rho^2+(z\pm i\mu)^2} \equiv X\pm iY,
\ee
with
\be             \label{plus}
X=\pm {\cal X},~~~~~Y=\pm\sign(z)\,{\cal Y}, 
\ee
where 
\be                         \label{XY}
{\cal X}=\frac{1}{\sqrt{2}}\sqrt{
\sqrt{ 
A^2+B^2}
+A
}\,    ,~~~~~~~~~~~~~
{\cal Y}=\frac{1}{\sqrt{2}}\sqrt{
\sqrt{ A^2+B^2}
-A
}  \,,  
\ee
with 
$A=\rho^2+z^2-\m^2$ and  $B=2\m z$.
Since $R$ is replaced by a real quantity, 
\be
R=\frac12(R_{+}+R_{-})\to
X, 
\ee
and one has 
\be
\frac12\ln\left(
\frac{R-m}{R+m}
\right)
\to
\frac12\ln\left(
\frac{X-i\mu }{X+i\mu}
\right)
=-i\arctan\left(
\frac{X}{\mu}
\right),
\ee
it follows that the solution remains real-valued,
\be                             \label{WWW}
U= \sigma \arctan\left(
\frac{X}{\mu}
\right),~~~~~~~~~
k= \frac{\sigma^2}{2}\ln\left(
\frac{X^2+Y^2}
{X^2+\m^2}\right). 
\ee
These are  the  {\it oblate} Zipoy-Voorhees solutions \cite{Zipoy,Voorhees:1971wh}. 
These solutions are relatively little known, therefore we shall describe some of their properties.

The two sign choices $``\pm"$ in \eqref{plus} 
correspond to the two square root branches. For each branch the real and imaginary parts of 
the square root,  $X$ and $Y$, are 
discontinuous and/or not smooth at the branch cut along the segment \eqref{opt1a}. 
However, gluing the two branch sheets together gives a Riemann surface on which all functions are 
smooth and continuous. The functions $X$, $Y$ defined by \eqref{plus} are already 
smooth and continuous 
for $\rho> \mu$, while for $\rho<\mu$ one should analytically continue them through the cut, 
which is achieved by replacing
\be                \label{anal}
X\to \sign(z) X=\pm \sign(z){\cal X},~~~~~~Y\to \sign(z) Y=\pm {\cal Y}.
\ee
These functions are continuous for $\rho<\mu$ (although discontinuous for $\rho> \mu$). 
In particular, at the symmetry axis one has after the analytic continuation 
either $X=z$, $Y=\m$ or $X=-z$, $Y=-\mu$. 

\begin{figure}[th]
\hbox to \linewidth{ \hss

	\resizebox{7cm}{5cm}{\includegraphics{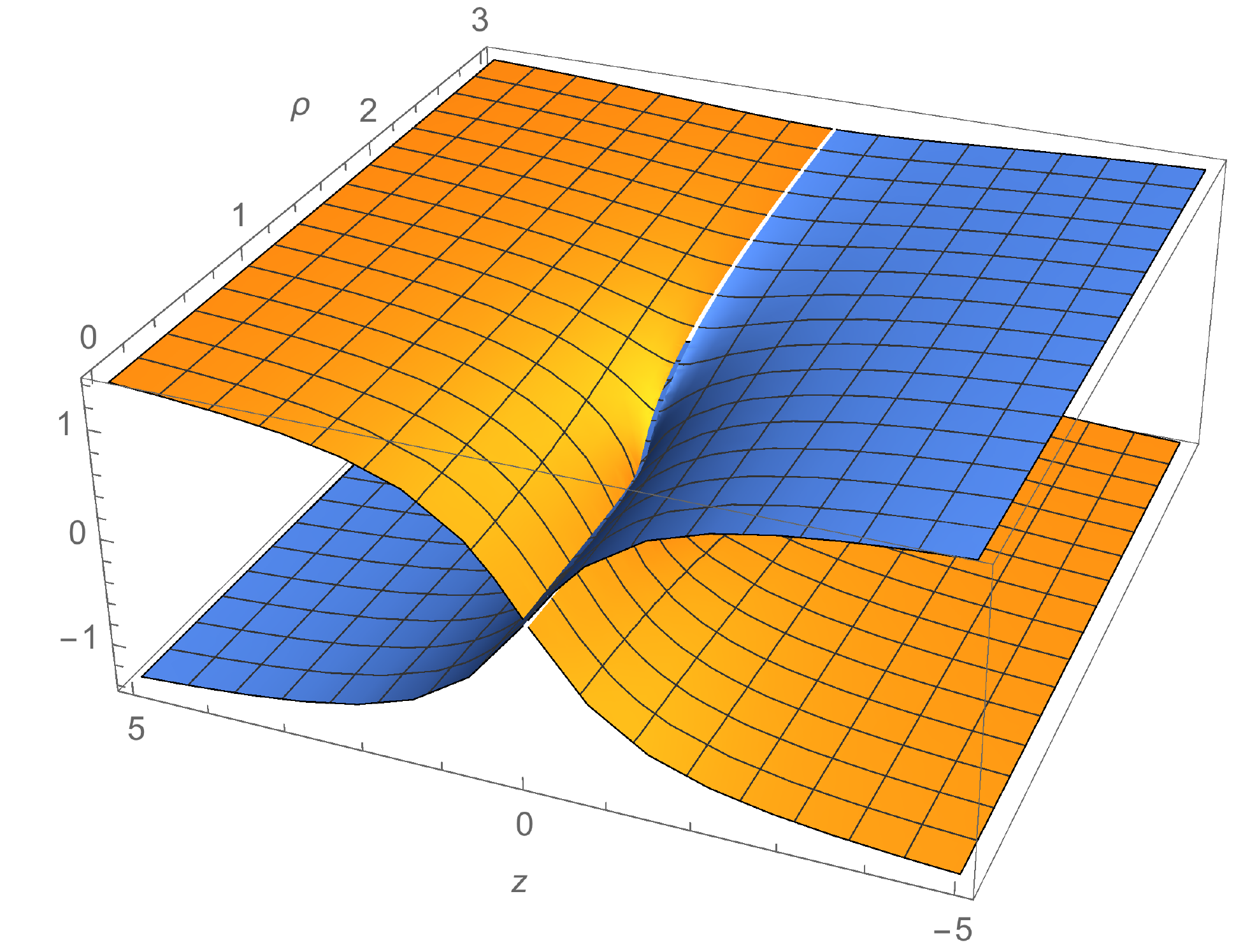}}
\hspace{1mm}
	
	\resizebox{7cm}{5cm}{\includegraphics{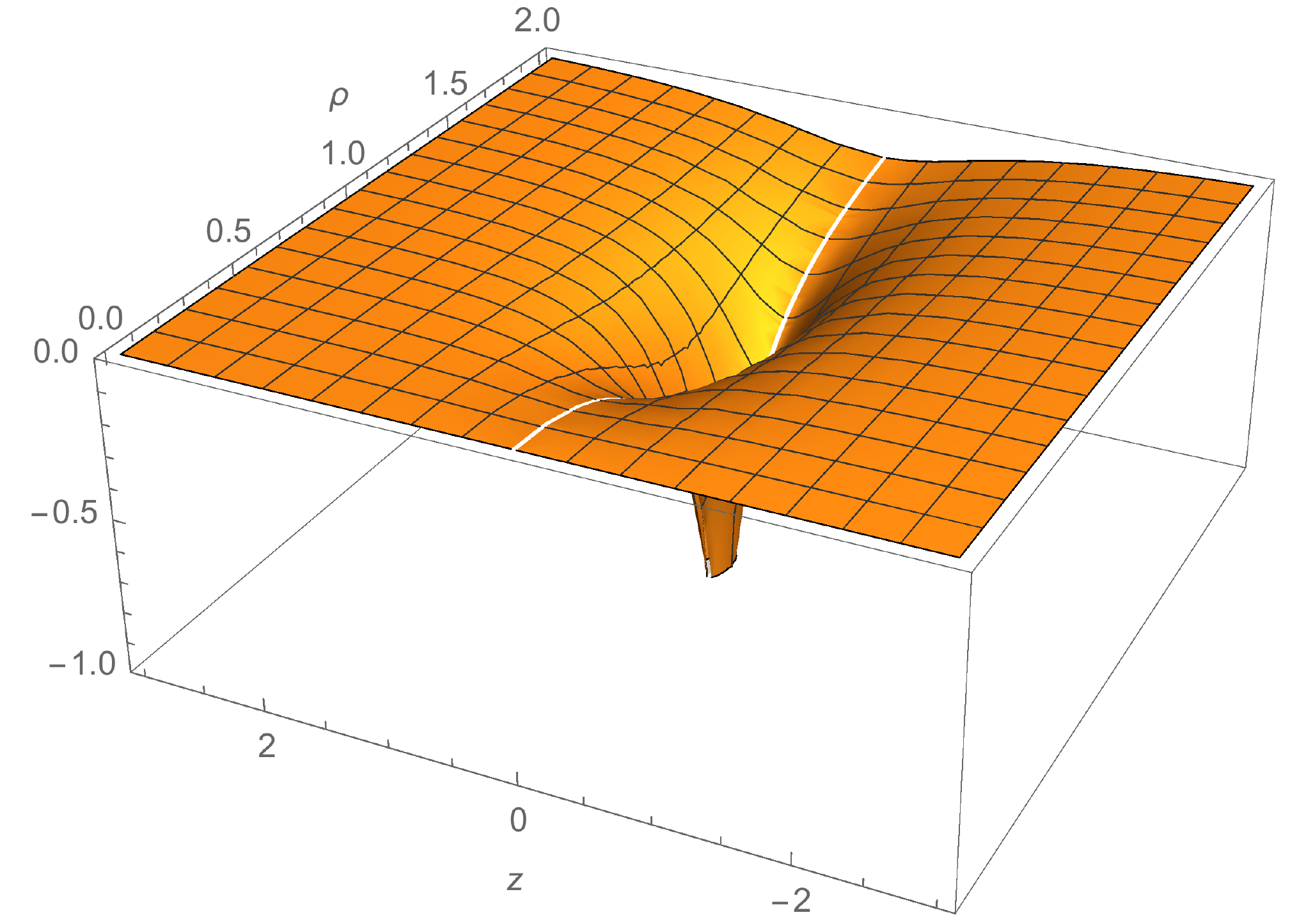}}
	
\hspace{1mm}
\hss}
\caption{$U(\rho,z)$ (left) and $k(\rho,z)$ (right) for the solution \eqref{WWW} with $\sigma=\mu=1$. 
The two branches of $U$ (blue and yellow online) correspond to $X$ chosen according \eqref{anal}.}
\label{FigUk}

\end{figure}

As a result, one obtains the plots of the solution \eqref{WWW} shown in Fig.\ref{FigUk}.
The function $U(\rho,z)$ is double-valued while $k(\rho,z)$ is single-valued and vanishes 
at the symmetry axis. The two different values of $U$ for given $\rho,z$ correspond to two different spacetime 
points.

\subsubsection{Wormhole topology}

 As first noticed by Zipoy  \cite{Zipoy}, 
solutions \eqref{WWW} describe wormholes. 
This can be seen by making the complexification 
\eqref{an} directly in \eqref{ZV}, which gives 
\bea                        \label{ZV1}
ds^2&=&-e^{2U} dt^2+e^{-2U} dl^2,  ~~~~~~~~~~~~~U=\sigma \arctan\left(\frac{\x}{\m}\right), \\
dl^2&=&\left(
\frac{\x^2+\m^2\cos^2\vartheta  }{x^2+\m^2  }
\right)^{1+\sigma^2}
\left[d\x^2+(\x^2+\m^2) d\vartheta^2\right]+(x^2+\m^2)\sin^2\vartheta d\varphi^2\,.  \nonumber 
\eea
This is the standard form of the  {\it oblate} Zipoy-Voorhees solutions \cite{Zipoy,Voorhees:1971wh}. 
When written in this form, it is seen that these 
metrics describe wormholes  with two asymptotically flat for $\x\to\pm\infty$ regions  
connected by a throat at $\x=0$. This
is  obvious already by noticing that close to 
the symmetry axis where $\cos^2\vartheta\approx 1$  the metric \eqref{ZV1} reduces to the 
wormhole metric \eqref{W} (up to the replacement $\sigma\to 1/s$).

The coordinates $(\x,\vartheta)$ are global and all their functions are single-valued,
while the 
Weyl coordinates are not global and one needs two Weyl charts  \cite{Clement:1983tu}
\be                    \label{Weylcharts}
D_\pm:\{\rho\geq 0,\, -\infty<z<\infty \}
\ee
to cover the manifold.
Let us 
discuss the relation between the two coordinate systems.

\begin{figure}[th]
\hbox to \linewidth{ \hss

	\resizebox{11cm}{9cm}{\includegraphics{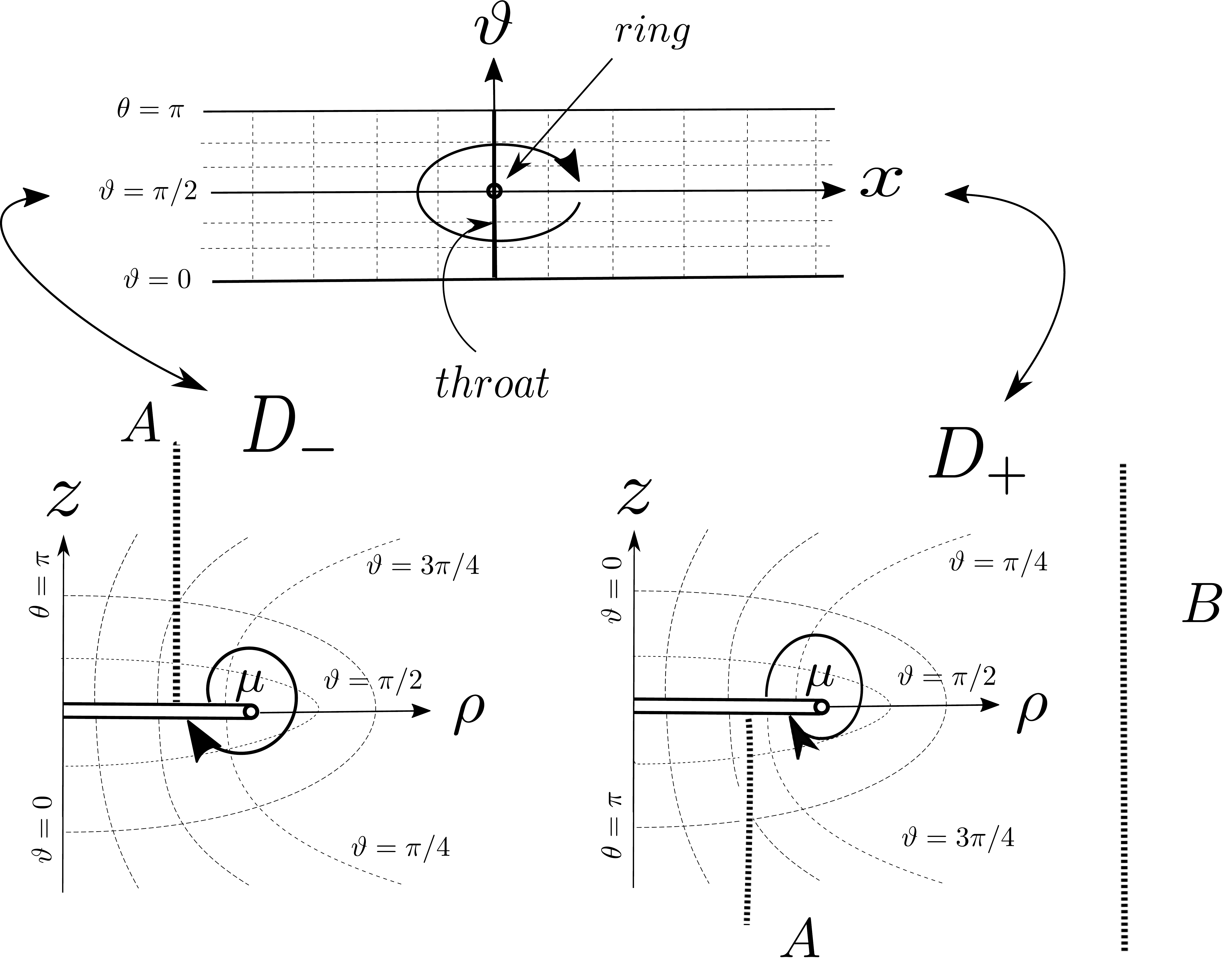}}
\hspace{1mm}
	
\hspace{1mm}
\hss}
\caption{Wormhole topology: The $x,\vartheta$ coordinates cover the whole of the manifold, while
each of the two 
 Weyl charts $D_{+}$ and $D_{-}$ covers only either the $x>0$ or $x<0$  part of the manifold.
The upper edge of the cut in the $D_{+}$ patch is identified with the lower edge of the $D_{-}$ cut
and vice-versa.  A winding around the ring in the  $x,\vartheta$ coordinates corresponds to two windings
in Weyl coordinates. }
\label{Fig}

\end{figure}

The $(\rho,z)$ coordinates are related to $(x,\vartheta)$ via \eqref{x2},
\be               \label{zzz}
z=x\cos\vartheta,~~~~\rho=\sqrt{x^2+\m^2}\,\sin\vartheta,
\ee
therefore, 
\be            \label{zzz1}
X\pm iY=
\sqrt{\rho^2+(z\pm i\mu)^2} =\sqrt{(x\pm i\m\cos\vartheta)^2}=x\pm i\m\cos\vartheta,
\ee
hence
\be            \label{zzz2}
x=X(\rho,z),~~~~~~\m\cos\vartheta=Y(\rho,z), 
\ee
which gives  inverse transformation from $(\rho,z)$ to $(x,\vartheta)$.
Since 
for given $(\rho,z)$ there are two different branches of 
$(X,Y)$ defined by \eqref{plus},
the inverse transformation is double-valued -- 
the same pair of values $(\rho,z)$ 
can correspond either to $(x,\vartheta)$ or to $(-x,\pi-\vartheta)$. 
Therefore, the $D_{+}$ chart covers the ``positive" ($x>0$) wormhole side
while  the $D_{-}$ chart covers  the ``negative"  ($x<0$)  side (see Fig.\ref{Fig}) \cite{Egorov:2016rfr}. 

The symmetry axis in Weyl coordinates, $\rho=0$, $z\in(-\infty,+\infty)$,
has two images in the spacetime: either $\vartheta=0$, $z=x$ or 
$\vartheta=\pi$, $z=-x$. Therefore, the spacetime actually has two symmetry axes, one at $\vartheta=0$ and 
one at $\vartheta=\pi$, 
corresponding to the edges of the $(x,\vartheta)$ strip in Fig.\ref{Fig}.
This explains, in particular,  the two values of $U(\rho,z)$ in Fig.\ref{FigUk}. 
For example, at $\rho=0$ one has for the (yellow online in  Fig.\ref{FigUk}) branch of $U(\rho,z)$ with 
 $X=x=z$ and $Y=\mu$ 
\be                      \label{v1}
U(0,z)=\sigma \arctan\left(\frac{z}{\mu}\right)=\sigma \arctan\left(\frac{x}{\mu}\right).
\ee
This gives the value at the $\vartheta=0$ axis. 
Similarly, for the second (blue online) branch 
of $U$ in  Fig.\ref{FigUk} one has 
 $X=x=-z$ and $Y=-\mu$ at $\rho=0$, hence 
\be                      \label{v2}
U(0,z)=-\sigma \arctan\left(\frac{z}{\mu}\right) =\sigma \arctan\left(\frac{x}{\mu}\right), 
\ee
which gives the value at the $\vartheta=\pi$ axis. These two values of $U(0,z)$ 
coincide  when expressed in terms of the $x$-coordinate.

It is also instructive to pass to the Weyl coordinates directly in \eqref{ZV1}. 
The starting point is the line element 
\be
dl^2&=&d\x^2+(\x^2+\m^2) (d\vartheta^2+\sin^2\vartheta d\varphi^2)
\ee
where $x\in(-\infty,\infty)$. 
This can be transformed to the isotropic form by setting 
\be
x=\frac{\xx^2-\m^2}{2\xx}~~~~\Leftrightarrow~~~~\xx=\x+ \sqrt{\x^2+\m^2}\in [0,\infty), 
\ee
which gives 
\be
dl^2 =\frac14\left(1+\frac{\m^2}{\xx^2}\right)^2[d\xx^2+\xx^2 d\vartheta^2+\xx^2\sin^2\vartheta d\varphi^2].
\ee
Introducing the complex variable 
\be
\omega=i\xx e^{-i\vartheta}=\xx\sin\vartheta+i\xx\cos\vartheta
\ee
one obtains 
\be
dl^2 =\frac14\left(1+\frac{\m^2}{|\omega|^2}\right)^2[|d\omega|^2+(\Re(\omega))^2d\varphi^2].
\ee
The $\omega$-variable sweeps the upper half-plane
$\Re(\omega)\geq 0$ such that the $x<0$ and $x>0$ regions map, respectively, to the 
$|\omega|<\m$ and  $|\omega|>\m$ regions. 
Passing to the new complex variable $\omega\to w(\omega)$
via the Joukowski transformation  \cite{Clement:1983ic}, 
\be                  \label{J}
w=\frac12\left(\omega+\frac{\m^2}{\omega}\right)\equiv \rho+iz,
\ee
one finds  that $\rho=\Re(w)$ and $z=\Im(w)$ are related to $x,\vartheta$ by \eqref{zzz},
whereas 
the metric becomes 
\be
dl^2&=&\frac14\left(1+\frac{\m^2}{|\omega|^2}\right)^2 \left|\frac{d\omega}{dw}\right|^2|dw|^2+\rho^2 d\varphi^2.
\ee
As a result, one has 
\bea
dx^2+(x^2+\m^2)d\vartheta^2&=&\frac14\left(1+\frac{\m^2}{|\omega|^2}\right)^2 \left|\frac{d\omega}{dw}\right|^2|dw|^2
=\frac{\x^2+\m^2}{\x^2+\m^2\cos^2\vartheta} (dz^2+d\rho^2), \nonumber \\
(x^2+\mu^2)\sin^2\vartheta d\varphi^2&=&\rho^2d\varphi^2,
\eea
using which and assuming \eqref{zzz2},  the solution \eqref{ZV1} takes up the Weyl
form with $U,k$ given by \eqref{WWW}. 

\begin{figure}[th]
\hbox to \linewidth{ \hss

	\resizebox{10cm}{7cm}{\includegraphics{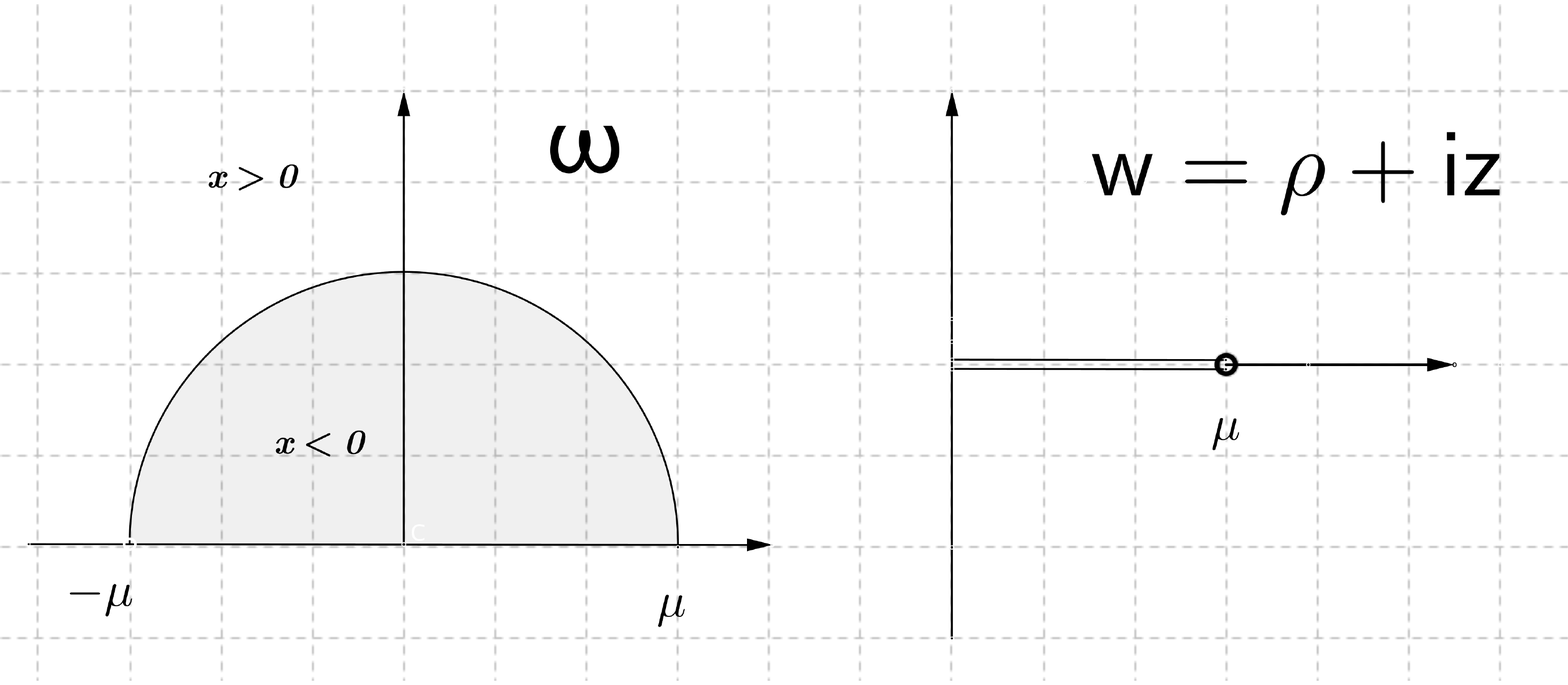}}
\hss}
\caption{The Joukowski  function \eqref{J} maps the interior $(x<0)$ 
of the semi-cercle $|\omega|=\mu$, $\Im(\omega)\geq 0$
to a half-plane $\Re(w)\geq 0$ with a cut removed, which corresponds to the $D^{-}$ Weyl region. 
The exterior $(x>0)$ of the semi-circle maps similarly to the $D^{+}$ Weyl region. 
}
\label{FigJ}

\end{figure}

The Joukowski transformation $\omega\to w$ \eqref{J}
 maps  the regions $|\omega| <\mu$
and $|\omega| >\mu$ 
of the $\omega$-plane  to the 
whole $w$-plane with the removed segment or real axis, 
\be                         \label{seg}
w\in[-\mu, \mu]. 
\ee
The inverse Joukowski transformation 
\be              \label{omega}
\omega=w\pm\sqrt{w^2-\mu^2}
\ee
has branching points at $w=\pm \mu$ connected by the branch cut \eqref{seg}. Any closed contour 
intersecting the cut passes from one Riemann sheet to the other and the sign of the square root in \eqref{omega}
changes. As a result, the full Riemann surface on which the Joukowski function is single-valued 
consistes of two sheets glued together through
the cut \eqref{seg}.

Restricting now to the upper half-plane $\Im(\omega)\geq 0$, the ``positive" ($|\omega|>\mu$) and ``negative" 
($|\omega|<\mu$)
wormhole sides map, respectively, to the $D_{+}$ and  $D_{-}$ Weyl regions \eqref{Weylcharts},
each region having the branch cut \eqref{opt1a} removed (see Fig.\ref{FigJ}).  
 The entire coordinate 
atlas of the spacetime manifold  is obtained  by identifying the upper edge of the cut 
on the $D_{+}$ chart with the lower edge of the cut on the $D_{-}$ chart 
and vice-versa. The negative part of the $z$-axis in the $D_{-}$ patch 
then merges with the positive part of the $z$-axis in the $D_{+}$ patch (see Fig.\ref{Fig}), which gives 
the $\vartheta=0$ axis; similarly for the $\vartheta=\pi$ axis. 

It is worth emphasising  again that the wormhole features of the  oblate vacuum solutions 
had been recognised  already in 1966 by Zipoy  \cite{Zipoy}, but even now 
these solutions remain very little known.

  \subsubsection{Ring wormholes   \label{rwh}}
 
The branching point of the Weyl coordinates at $\rho+iz=\mu$ gives rise to a curvature singularity 
at $\x=0$, $\vartheta=\pi/2$,  that is along  a circle  of radius $\m$ placed in the 
equatorial plane. The curvature consists of the Ricci part 
and of the Weyl  part, which show, respectively, 
a distributional singularity and a power-law singularity. 
Consider first the distributional part. This arises because the Ricci tensor vanishes everywhere 
(the metric is vacuum) apart form the ring where it shows 
 a delta-type singularity. 
 
Setting $x=\mu\, x_1$ and $\cos\vartheta=x_2$,
the metric  \eqref{ZV1} becomes
\bea                        \label{ZV2}
ds^2&=&-e^{2U} dt^2+ e^{-2U} dl^2,  ~~~~~~~~~~~~~U=\sigma \arctan\left(x_1\right), \\
dl^2&=&\mu^2\left(
\frac{x_1^2+x_2^2  }{x_1^2+1  }
\right)^{1+\sigma^2}
\left[dx_1^2+\frac{x_1^2+1}{1-x_2^2}\, dx_2^2\right]+\mu^2(x_1^2+1)(1-x_2^2)\, d\varphi^2\,.  \nonumber 
\eea
Expanding this for small $x_1,x_2$ yields in the leading order 
\be                  \label{XPEH}
ds^2 = -dt^2+\mu^2 \left(
{x_1^2+x_2^2}
\right)^{1+\sigma^2}(dx_1^2+dx_2^2)+\mu^2d\varphi^2+\ldots ,
\ee
where the dots denote  the subleading terms. 
Passing to polar coordinates $r,\theta$ via
\be
(x_1+ix_2)^q=\frac{q}{\mu}\,{r}e^{i\theta}
\ee
with $q=2+\sigma^2$
yields 
\be                    \label{met}
ds^2= -dt^2+dr^2+r^2 d\theta^2+\mu^2d\varphi^2+\ldots
\ee
This metric looks flat, but since 
the angular variable $\theta$ ranges from zero to $2\pi q>2\pi$ after a revolution around the point  $r=0$, 
there is %
a negative angle deficit, 
\be
\delta=2\pi-2\pi q=-(\sigma^2+1)2\pi,
\ee
which gives rise to a conical singularity of the Ricci tensor $r=0$. 
This can be associated to  a distributional matter source -- 
an infinitely thin ring of radius $\mu$ and 
of {\it negative} tension  (energy per unit length) \cite{Gibbons:2016bok} 
\be
T=-\frac{(1+\sigma^2)c^4}{4G}
\ee 
(the correct physical dimensions are restored here for the moment).
The  ring encircles the wormhole throat and plays the role of the negative energy source for the solution. 

Consider now the power-law singularity of the Riemann tensor. 
Although the leading terms of the 
 line element in \eqref{met} describe a  (locally) flat geometry, 
the subleading terms denoted by the dots  produce a non-zero curvature. 
Since these terms vanish for $x_1,x_2\to 0$, one could expect the curvature to vanish in this limit too. 
Surprisingly, just the opposite happens and the curvature diverges for $x_1,x_2\to 0$ 
because  the subleading terms do not vanish fast enough in this limit. 
To understand this, consider just one metric component, 
\be
g_{\varphi\varphi}=\mu^2 \exp[-2\sigma\arctan(x_1)](x_1^2+1)(1-x_2^2)=\mu^2( 1+\ldots ).
\ee 
Keeping only the leading order term here would give a constant, $\mu^2$,  whose derivatives vanish and
give no contribution to the curvature. However, keeping also the subleading terms,
the second derivatives $\partial_{11}g_{\varphi\varphi}$ and $\partial_{22}g_{\varphi\varphi}$ 
and the $R_{1\varphi 1\varphi}$ and  $R_{2\varphi 2\varphi}$ curvature components do not vanish
at $x_1,x_2\to 0$, implying 
that 
 the curvature invariant diverges. A direct computation  yields 
\be                  \label{curv}
R_{\mu\nu\alpha\beta}R^{\mu\nu\alpha\beta}=
\frac{\sigma^2}{\mu^4}\frac{16(1+\sigma^2)^2}{(x_1^2+x_2^2)^{3+2\sigma^2} }\,
(1+\ldots )
\ee
(the dots denoting terms that vanish for $x_1,x_2\to 0$)  hence   the curvature 
is unbounded for small $x_1,x_2$. This shows that the form of the line element \eqref{met} is 
somewhat deceptive  -- even though close to the ring it manifestly approaches flat metric (with an angle deficit), 
the curvature grows. 

Summarising,  the ring carries the distributional conical singularity and also the power-law
singularity of the curvature. 

Notice that the curvature tends to zero if $\mu\to\infty$. This is natural, since in this limit the ring radius tends to infinity.
 Each its finite segment then approaches a piece of an infinite   straight cosmic string whose geometry 
is locally flat.  Therefore, one can view the Weyl curvature of the ring  as an effect of  bending the cosmic string.  

Another remarkable limit where the geometry becomes locally flat is obtained by taking $\sigma\to 0$. 
The curvature then vanishes since 
the functions $U,k$ in \eqref{WWW}  vanish  and
the metric expressed in Weyl coordinates becomes manifestly flat,
\be                         \label{flat}
ds^2=-dt^2+d\rho^2+dz^2+\rho^2 d\varphi^2.
\ee
However, the topology is still non-trivial because the $(\rho,z)$ coordinates  do not cover the 
whole of the manifold but only either the $D_{+}$ or $D_{-}$ patches glued to each other through the cuts. 
A contour around the ring core 
in the $D_{+}$ patch does not close after a revolution of $2\pi$ but passes to the $D_{-}$ patch and only
after a second revolution of $2\pi$ returns back to the initial position (see Fig.\ref{Fig}). 
Therefore, the winding angle around the ring core 
ranges from zero to $4\pi$, which indicates that the ring is still there and has   
the tension 
\be              \label{tens}
T=-\frac{c^4}{4G}.
\ee 
As a result, the ring supports only the conical singularity while the geometry is locally exactly flat. 

\begin{figure}[th]
\hbox to \linewidth{ \hss

	\resizebox{7cm}{4cm}{\includegraphics{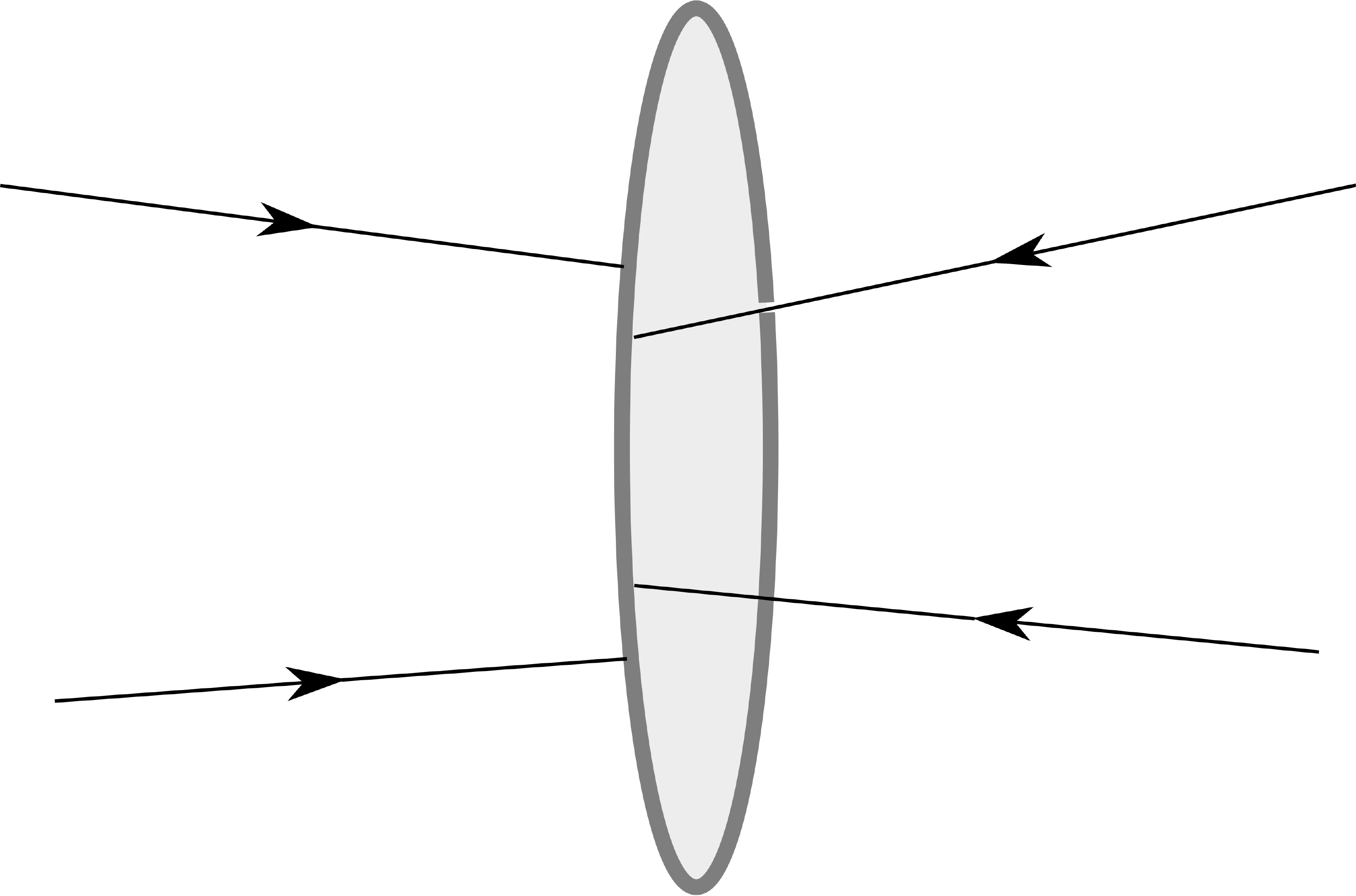}}
\hspace{1mm}
	
	
\hspace{1mm}
\hss}
\caption{Particles entering the ring are not seen coming out from the other side}
\label{Figf}

\end{figure}

It follows that the geodesics which do not hit the ring are simply straight  lines as in flat space.  
Those which pass outside the ring always stay in the same coordinate chart (line B in Fig.\ref{Fig}), 
while those threading  the ring pass to the other chart and become invisible 
from the previous chart thus traversing the wormhole (line A in Fig.\ref{Fig}). 
Therefore, the ring genuinely cuts  a hole in  flat space through which on can observe another universe 
and get there. This reminds one of Alice observing the room behind the looking glass and next jumping there. 
An object falling through the ring can be seen from behind, while viewed from the side 
it is not seen coming from the other side (see Fig.\ref{Figf}). 

It seems plausible that the wormhole edges could be smoothened without changing 
the global structure if the singular ring is replaced by a regular hoop-shaped
energy distribution  of finite  thickness and with the same tension. 
Inside the hoop the energy density is finite hence the 
geometry must be regular, while outside the energy is zero and the geometry 
should be more or less the same 
as for the original ring. 
This suggests that wormholes could be created by negative energies concentrated  in toroidal volumes, 
for example by vacuum fluctuations.
Therefore, traversable wormholes could 
actually be more than just a mathematical curiosity.

However,  the energy needed to create a ring wormhole 
is very high. 
The absolute value of the negative 
tension $T$ in \eqref{tens} coincides with the highest possible value for a {\it positive} tension
(force), according to 
the maximum tension principle in General Relativity conjectured in  \cite{Schiller:2006cm,Gibbons:2002iv}. 
This conjecture is supported, for example,  
by the fact that the angle deficit of a cosmic string cannot exceed $2\pi$. 
Numerically one has ${T}=-3.0257\times 10^{43}$~Newtons~$\approx -3\times 10^{39}$ Tonnes, 
hence 
to create a macroscopic 
ring wormhole of radius $R=1$ metre, say, one needs a negative energy equivalent to the mass of Jupiter, 
$2\pi RT/c^2\approx 10^{-3} M_\odot$ \cite{Gibbons:2016bok,Krasnikov:2002dq}. 

At the same time, 
one can also imagine microscopically small  rings by quantum fluctuations 
appearing spontaneously from the vacuum and then disappearing again. Particles crossing the 
ring during its existence would no longer be accessible from our universe after the ring disappears. 
If true, this would be a potential 
mechanism for the loss of quantum coherence. 

It is worth noting that the $\sigma=0$ ring wormhole with locally flat geometry can be viewed 
as a particular case of the so called   ``star gate  wormholes" constructed 
by surgeries and identifications performed on  Minkowski space \cite{Visser:1995cc}. In such a construction two copies of
Minkowski space are identified over a boundary of a compact 3-region removed from each copy
and the geometric junction conditions  determine  the surface 
energy density, which  is negative for convex surfaces. 
In the simplest cases such wormholes are ``cubic" or  ``polyhedral" 
with the negative energy localized only along the polyhedra edges 
\cite{Visser:1989kh}.  Their particular limit are planar ``dihedral wormholes", 
which in their  turn reduce to ``loop based wormholes" when the number of edges tends to infinity \cite{Visser:1995cc}.
For such wormholes the negative energy is distributed along a loop of arbitrary shape encircling the wormhole throat,
the geometry being exactly flat outside the loop. Remarkably, the loop tension 
turns out to be  
precisely the same as the value \eqref{tens} obtained in our analysis by completely different methods. 
As a result, our estimate of the negative energy equivalent to the mass of Jupiter 
needed to create a one meter size wormhole is actually the same as the one previously  quoted
in the literature \cite{Visser:1995cc}.

\subsection{Solutions with scalar field   \label{sec-scal}}

We can now produce new solutions with scalar field by 
applying the rotations  \eqref{sym1} and boosts \eqref{sym2} 
to the prolate and oblate vacuum metrics written in the form \eqref{ZV} and \eqref{ZV1}. 
Applying the rotation \eqref{sym1}  produces solutions with the conventional scalar 
in the prolate case, 
\bea                        \label{ZVI}
ds^2&=&-\left(
\frac{x-m}{x+m}
\right)^{\lambda/s} dt^2+\left(
\frac{x-m}{x+m}
\right)^{-\lambda/s} dl^2,  ~~~~\phi=\frac{\sqrt{s^2-1}}{s}\,\frac{\lambda}{2}\ln \left(
\frac{x-m}{x+m}
\right),  \nonumber   \\
dl^2&=&\left(
\frac{\x^2-m^2\cos^2\vartheta  }{x^2-m^2  }
\right)^{1-\lambda^2}
\left[d\x^2+(\x^2-m^2) d\vartheta^2\right]+(x^2-m^2)\sin^2\vartheta d\varphi^2 , 
\eea
and in the oblate case, 
\bea                        \label{ZVII}
ds^2&=&-e^{2(\sigma/s)\Psi} dt^2+e^{-2(\sigma/s)\Psi} dl^2, ~~~~~\phi=\sigma \frac{\sqrt{s^2-1}}{s}\,\Psi,  \\
dl^2&=&\left(
\frac{\x^2+\m^2\cos^2\vartheta  }{x^2+\m^2  }
\right)^{1+\sigma^2}
\left[d\x^2+(\x^2+\m^2) d\vartheta^2\right]+(x^2+\m^2)\sin^2\vartheta d\varphi^2 \nonumber, 
\eea
where $\Psi=\arctan(\x/\m)$ and $s\geq 1$. 
Applying boosts \eqref{sym2}  gives solutions with the phantom field -- 
boosted prolate solutions, 
\bea                        \label{ZVIII}
ds^2&=&-\left(
\frac{x-m}{x+m}
\right)^{\lambda/s} dt^2+\left(
\frac{x-m}{x+m}
\right)^{-\lambda/s} dl^2,  ~~~~\psi=\frac{\sqrt{1-s^2}}{s}\,\frac{\lambda}{2}\ln \left(
\frac{x-m}{x+m}
\right),   \nonumber   \\
dl^2&=&\left(
\frac{\x^2-m^2\cos^2\vartheta  }{x^2-m^2  }
\right)^{1-\lambda^2}
\left[d\x^2+(\x^2-m^2) d\vartheta^2\right]+(x^2-m^2)\sin^2\vartheta d\varphi^2,
\eea
and boosted oblate solutions, 
\bea                        \label{ZVIV}
ds^2&=&-e^{2(\sigma/s)\Psi} dt^2+e^{-2(\sigma/s)\Psi} dl^2, ~~~~~\psi=\sigma \frac{\sqrt{1-s^2}}{s}\,\Psi,  \\
dl^2&=&\left(
\frac{\x^2+\m^2\cos^2\vartheta  }{x^2+\m^2  }
\right)^{1+\sigma^2}
\left[d\x^2+(\x^2+\m^2) d\vartheta^2\right]+(x^2+\m^2)\sin^2\vartheta d\varphi^2,  \nonumber 
\eea
where $\Psi$ is the same as before and $s\leq 1$. 
Finally, new solutions with the phantom scalar can be obtained by applying the swap symmetry 
$(U,k,\psi)\to (\psi,-k,U)$. This gives the prolate solutions, 
\bea                        \label{ZVV}
ds^2&=&-\left(
\frac{x-m}{x+m}
\right)^{\lambda/s} dt^2+\left(
\frac{x-m}{x+m}
\right)^{-\lambda/s} dl^2,  ~~~~\psi=\frac{\sqrt{1+s^2}}{s}\,\frac{\lambda}{2}\ln \left(
\frac{x-m}{x+m}
\right),   \nonumber   \\
dl^2&=&\left(
\frac{\x^2-m^2\cos^2\vartheta  }{x^2-m^2  }
\right)^{1+\lambda^2}
\left[d\x^2+(\x^2-m^2) d\vartheta^2\right]+(x^2-m^2)\sin^2\vartheta d\varphi^2,
\eea
and the oblate solutions, 
\bea                        \label{ZVVI}
ds^2&=&-e^{2(\sigma/s)\Psi} dt^2+e^{-2(\sigma/s)\Psi} dl^2, ~~~~~\psi=\sigma \frac{\sqrt{1+s^2}}{s}\,\Psi,  \\
dl^2&=&\left(
\frac{\x^2+\m^2\cos^2\vartheta  }{x^2+\m^2  }
\right)^{1-\sigma^2}
\left[d\x^2+(\x^2+\m^2) d\vartheta^2\right]+(x^2+\m^2)\sin^2\vartheta d\varphi^2,  \nonumber 
\eea
where now $s\in(-\infty,+\infty)$. The latter solutions reduce to the BE wormholes for $\sigma=1$. 

We have therefore obtained six families of new solutions, each carrying   three free parameters --
either $m,\lambda,s$ or $\m,\sigma,s$. The oblate solutions can be obtained from the prolate ones 
by analytic continuation  $m\to i\m$ and $\lambda\to -i\sigma$. We also notice that \eqref{ZVV},\eqref{ZVVI}
can  be obtained either from \eqref{ZVI},\eqref{ZVII} via $m\to i\m$, $s\to is$ or 
from \eqref{ZVIII},\eqref{ZVIV} via $\lambda\to i\lambda$, $s\to is$. 
As a result, further analytic continuations do not give new solutions. 

However, any of the above solutions can in addition be acted upon by the tachyon symmetry \eqref{tachyon} 
interchanging $t$ and $\varphi$.   This doubles the size of the family of new solutions. 
Most of these solutions are only axially symmetric and have never been described before. 
Here are some of their properties. 

\begin{itemize}
\item 
There are twice as many solutions with the phantom field as those with the conventional scalar --
owing to the swap symmetry $(U,k,\psi)\to (\psi,-k,U)$. 

\item
The rotated prolate solutions \eqref{ZVI} for the conventional scalar 
generalize the FJRW family \eqref{F} and reduce to it for $\lambda^2=1$. 
\item
The boosted prolate solutions  \eqref{ZVIII} for the phantom field reduce for $\lambda^2=1$ to the 
``phantom version" \eqref{FF} of the FJRW solutions.

\item
The (boosted and swapped) oblate solutions \eqref{ZVVI} for the phantom field reduce for $\sigma^2=1$ 
to the BE spherically symmetric wormholes \eqref{W}. For $\sigma^2\neq 0$ they describe 
axially symmetric deformations of the BE wormholes.

\item
The oblate solutions \eqref{ZVII},\eqref{ZVIV} 
describe ring wormholes supporting either conventional or phantom scalars.
When  $\sigma\to 0$, the scalar vanishes and the solutions 
reduce to the ring wormhole \eqref{flat} with locally flat geometry
$
ds^2=-dt^2+dl^2
$
%
where 
\bea                                 \label{flat0}
dl^2 &=& 
\frac{\x^2+\m^2\cos^2\vartheta  }{x^2+\m^2  }
\left[d\x^2+(\x^2+\m^2) d\vartheta^2\right]+(x^2+\m^2)\sin^2\vartheta d\varphi^2  \nonumber  \\
&=&d\rho^2+dz^2+\rho^2d\varphi^2.  
\eea

\item
The role of the prolate solutions \eqref{ZVV} is less clear -- they never reduce to the 
Schwarzschild metric although the geometry becomes flat as $\lambda\to 0$.

\item 
Solutions \eqref{ZVI}, \eqref{ZVII}, \eqref{ZVV}, \eqref{ZVVI} become ultrastatic $(U=0)$ for  $s\to\infty$
but the scalar field remains non-trivial in this limit. 
The oblate solutions \eqref{ZVVI} interpolate in this limit 
between the spherically symmetric ultrastatic BE wormhole for $\sigma^2=1$ and 
the ring wormhole \eqref{flat0} for $\sigma=0$. 

\item
Yet one more interesting limit is obtained by taking in solutions with phantom \eqref{ZVIII}--\eqref{ZVVI}
$\lambda\to 0$, $\sigma\to 0$, $s\to 0$ with fixed 
$\lambda/s$ and $\sigma/s$. The three dimensional part of the metric $dl^2$ then reduces to that 
in \eqref{flat0} and hence  becomes flat, 
but since 
$
U=\psi\neq 0
$, 
the 4-geometry 
remains non-trivial. 
This is a particular case of the anti-gravitating solutions described in Section \eqref{antigr}. 

\item 
Expressing the locally flat $dl^2$ in the 
$(x,\vartheta)$ coordinates as in \eqref{flat0} gives 
more  ring-type anti-gravitating solutions: 
$
ds^2=-e^{2U}dt^2+e^{-2U} dl^2
$
with harmonic $U=\psi$, 
\be                                   \label{lap}
\frac{\partial}{\partial x}\left((x^2+\mu^2)\frac{\partial U}{\partial x}\right)
+\frac{1}{\sin\vartheta}\frac{\partial}{\partial \vartheta}\left(\sin\vartheta\, \frac{\partial U}{\partial \vartheta}\right)=0. 
\ee
Choosing $U=\arctan(x/\mu)$ corresponds to the described above solutions.
 Other  solutions of \eqref{lap} can be explicitly found \cite{Matos:2009au},
but they seem to be all singular.  
For example, choosing the Appell potential \cite{Appell},
\be                                           \label{Uap}
U=\frac{\cos\vartheta}{x^2+\mu^2\cos^2\vartheta},
\ee
yields the ring-type solution previously studied in the literature 
\cite{Miranda:2013gqa,Matos:2012gj},  but the Newtonian potential 
 diverges at the ring position, 
$x=\cos\vartheta=0$. As discussed below in Section \ref{apl}, this solution can also be 
obtained from the Appell ring \eqref{Curz2} via the infinite boost. 

\end{itemize}

\section{Solutions from two rods}

Let us now apply the scaling symmetry \eqref{scale} 
to the two-rod solution \eqref{2rod}. 
First of all we notice that this symmetry 
can be extended since the mass density of each rod can be 
rescaled independently. As a result, there can actually be two scale parameters $\lambda_1$ and $\lambda_2$ 
and rescaling  the solution  \eqref{2rod} gives a larger 
 family of vacuum metrics, 
\be                                  \label{2rodd}
U=\lambda_1 U_1+\lambda_2 U_2,~~~~~~~~
k=\lambda_1^2\, k_1+\lambda_2^2\, k_2+\lambda_1\lambda_2\, k_{12},
\ee
where ($a=1,2$)
\be            \label{TR1}
U_a&=&\frac{1}{2}\ln\left(
\frac{R_a-m_a}{R_a+m_a}
\right),~~~~~~~~k_a=\frac12\ln\left( \frac{(R_a)^2-(m_a)^2}{R_{a+}R_{a-}  } \right),~~~\nonumber \\
k_{12}&=&\frac12\ln\left(
\frac{ (R_{1+}R_{2-}+z_{1+}z_{2-}+\rho^2) (R_{1-}R_{2+}+z_{1-}z_{2+}+\rho^2) }{
(R_{1+}R_{2+}+z_{1+}z_{2+}+\rho^2)(R_{1-}R_{2-}+z_{1-}z_{2-}+\rho^2)   }
\right),
\ee
with
\be
z_{a\pm}=z-z_a\pm m_a,~~~~~~~~R_{a\pm}=\sqrt{\rho^2+(z_{a\pm})^2}\,,~~~~~R_a=\frac12(R_{a+}+R_{a-}). 
\ee
These ``prolate'' solutions depend on 6 parameters $\lambda_a,m_a,z_a$.

The ``oblate" solutions can be obtained by extending 
 to complex parameter values, 
\be
\lambda_a \to i\sigma_a,~~~~~~m_a\to i\m_a\,.
\ee
The amplitudes $R_{a\pm}$ in \eqref{TR1} then become 
\be                 \label{XXX}
R_{a\pm}\to \sqrt{\rho^2+(z_a^\pm)^2} \equiv X_a\pm iY_a ,~~~~~~~~~R_a\to X_a,
\ee
where
\be           \label{ZZZ}
z_a^\pm=z-z_a\pm i\m_a. 
\ee
Computing the square root gives 
\be             \label{prop}
X_a=s_X(a){\cal X}_a,~~~~~Y_a=s_Y(a){\cal Y}_a,
\ee
where $s_X(a)$ and $s_Y(a)$ take values $+1$ or $-1$ and will be specified below, while 
\bea                         \label{XYa}
{\cal X}_a&=&\frac{1}{\sqrt{2}}\sqrt{
\sqrt{ 
(A_a)^2+(B_a)^2}
+A_a
} \,,   ~~~
{\cal Y}_a=\frac{1}{\sqrt{2}}\sqrt{
\sqrt{ 
(A_a)^2+(B_a)^2}
-A_a
}  \,
\eea
with
\be
A_a=\rho^2+(z-z_a)^2-\mu_a^2\,,~~~~~B_a=2\m_a(z-z_a). 
\ee
This gives the oblate solutions, 
\be                                  \label{2w}
U=\sigma_1 U_1+\sigma_2 U_2,~~~~~~~~
k=\sigma_1^2\, k_1+\sigma_2^2\, k_2+\sigma_1\sigma_2\, k_{12},
\ee
where
\be                \label{2w1}
U_a=\arctan\left(
\frac{X_a}{\mu_a}
\right),~~~~~~
k_a=\frac{1}{2}\ln\left(
\frac{X_a^2+Y_a^2}{X_a^2+\m_a^2}
\right),
\ee
and
\bea             \label{2w2}
k_{12}
&=&\frac{1}{2}\ln\left(\left|
\frac
{ (X_1+iY_1)(X_2+iY_2)+z_1^{+}z_2^{+}+\rho^2 } 
{ (X_1+iY_1)(X_2-iY_2)+z_1^{+}z_2^{-}+\rho^2} 
\right|^2
\right). 
\eea   

\subsection{Multi wormholes}

We are mainly interested in the oblate 
solutions \eqref{2w} describing a pair of wormholes
(they can be generalized to the multi wormhole case as explained around Eq.\eqref{Nrod}), 
hence we shall describe  some of their properties.

\begin{figure}[th]
\hbox to \linewidth{ \hss

	\resizebox{12cm}{8cm}{\includegraphics{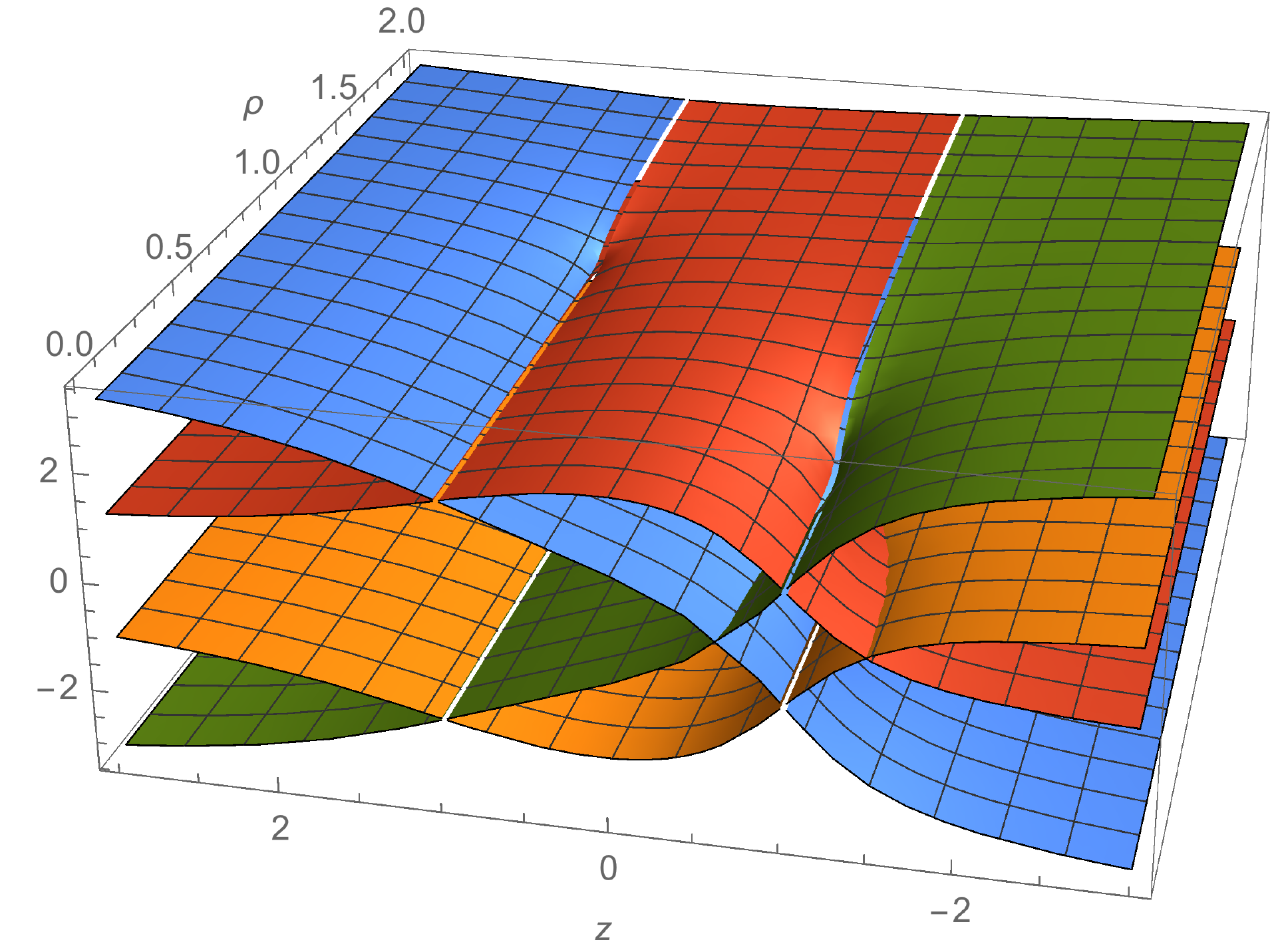}}
\hspace{1mm}
	
	
\hspace{1mm}
\hss}
\caption{$U(\rho,z)$ defined by  \eqref{2w} for $\sigma_1=1$, $\sigma_2=1.5$, 
$\mu_1=1.2$, $\mu_2=0.5$, $z_1=-z_2=1$, and for 
$\rho\in[0,2]$, $z\in[-4,4]$.
The four different branch sheets (different colours online) correspond to values of $U$ in 
four different spacetime regions. 
}
\label{FigU}

\end{figure}

\begin{figure}[th]
\hbox to \linewidth{ \hss

	\resizebox{8cm}{7cm}{\includegraphics{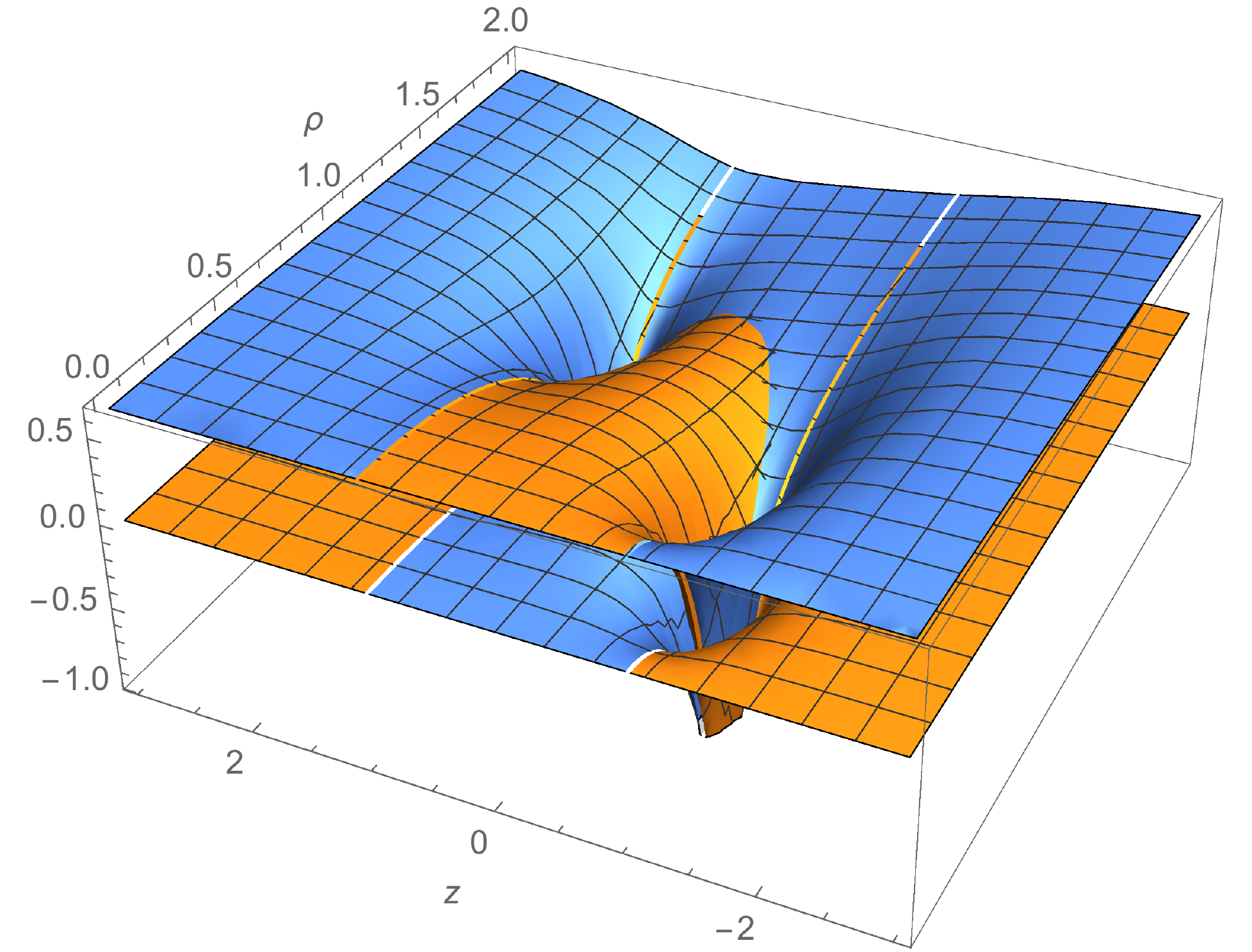}}
\hspace{1mm}
	
	\resizebox{8cm}{7cm}{\includegraphics{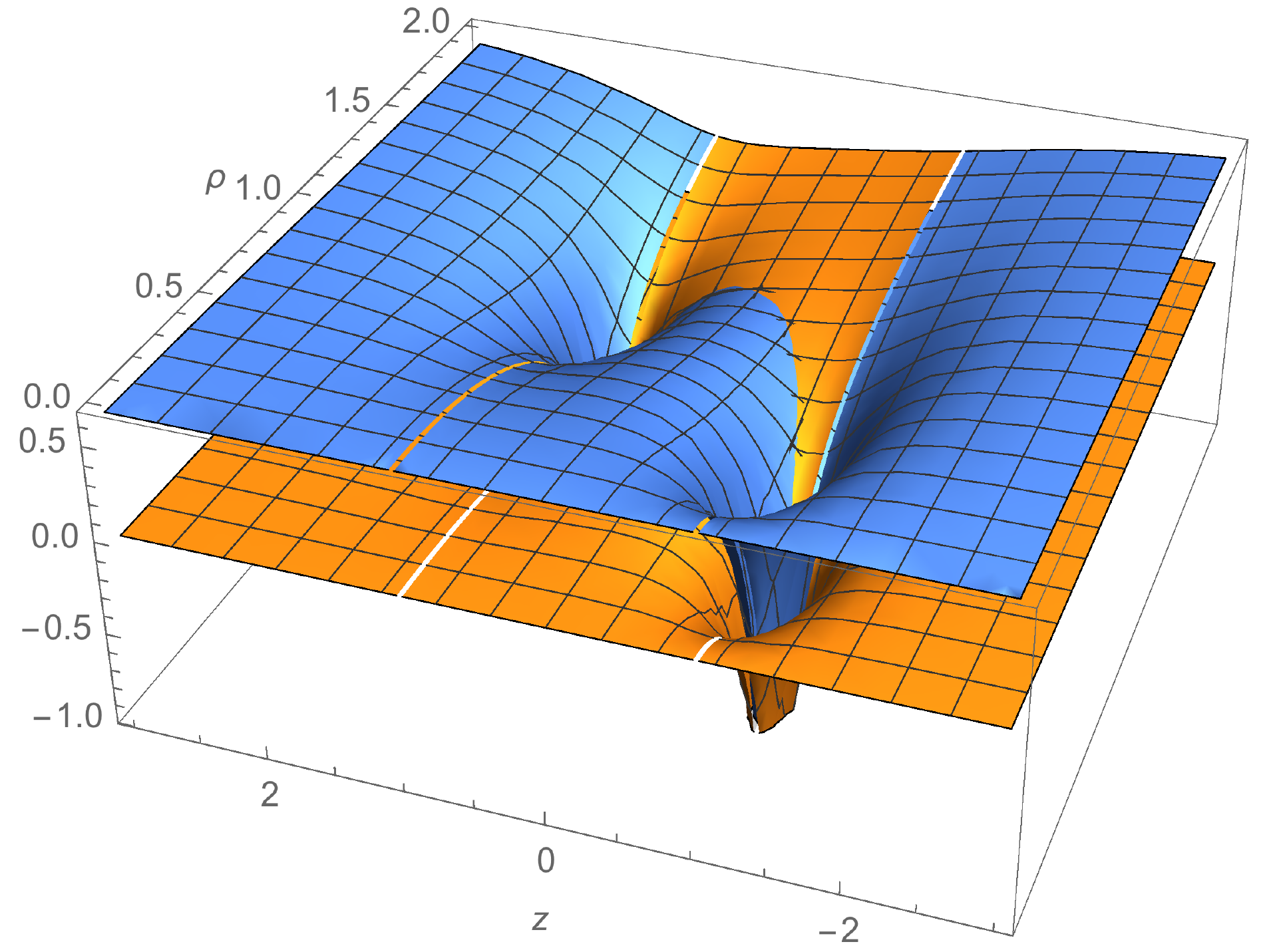}}
	
\hspace{1mm}
\hss}
\caption{$k(\rho,z)$ defined by \eqref{2w} for the same parameter values as in Fig.\ref{FigU}
and for $\rho\in[0,2]$, $z\in[-4,4]$. 
It shows only two sheets made of  (blue and yellow online) branches
computed from $X_a,Y_a$ chosen either as in \eqref{sign1} (left panel) or as in \eqref{sign2} (right panel). 
The resulting sheets are the same in both cases. 
}
\label{Figk}

\end{figure}

A closer inspection reveals that a proper choice of signs $s_X(a)$, $s_Y(a)$ in \eqref{prop}
is essential, since otherwise the field equations are fulfilled 
only in some regions of space. Specifically, 
since  $U_a$ and $k_a$ in \eqref{2w1} 
have the same structure as for the one wormhole solution, they satisfy 
\be                  \label{ee1}
\frac{\partial^2 U_{a}}{\partial\rho^2}&+&\frac{1}{\rho}\,\frac{\partial U_a}{\partial\rho}+\frac{\partial^2 U_a}{\partial z^2}=0, \nonumber \\
\frac{\partial k_a}{\partial\rho}&=&\rho\left[
\left(\frac{\partial U_a}{\partial\rho}\right)^2
-\left(\frac{\partial U_a}{\partial z}\right)^2
\right], \nonumber \\
\frac{\partial k_a}{\partial z}&=&2\rho\,
\frac{\partial U_a}{\partial\rho}
\frac{\partial U_a}{\partial z} , 
\ee 
and these equation hold everywhere. 
The  $k_{12}$ amplitude should satisfy 
\be                 \label{ee2}
\frac{\partial k_{12}}{\partial\rho}&=&2\rho\left[
\frac{\partial U_1}{\partial\rho}\frac{\partial U_2}{\partial\rho}
-\frac{\partial U_1}{\partial z}\frac{\partial U_2}{\partial z}
\right], \nonumber \\
\frac{\partial k_{12}}{\partial z}&=&2\rho\left[
\frac{\partial U_1}{\partial\rho}\frac{\partial U_2}{\partial z} +
\frac{\partial U_2}{\partial\rho}\frac{\partial U_1}{\partial z}
\right],
\ee 
but these equations do not always hold. For example, choosing in \eqref{prop} 
$s_X(a)=s_Y(a)=1$ these equations are satisfied  only if $z>z_1$ and $z>z_2$ 
and only if both $\m_1$ and $\m_2$ are positive. The reason is that $k_{12}$ 
depends on phases of $z_a^\pm$\,,
\be
\phi_a(z)=\arctan\left(\frac{\mu_a}{z-z_a} \right),
\ee
which jump when $z$ crosses values $z_a$. As a result, $k_{12}$ jumps too,
and the equations are satisfied only from one side of the jump. 
The proper sign choice for which  the equations are satisfied everywhere is 
\be                         \label{sign1}
X_a=s(a)\,{\cal X}_a,~~~~~~Y_a=s(a)\,\sign(\m_a)\,\sign(z-z_a)\,\,{\cal Y}_a,
\ee
which implies that the $Y_a$ amplitudes experience jumps through the cuts
\be            \label{cuts1}
{\cal I}_a=\{\rho\in[0,|\mu_a |],z=z_a\}.
\ee
There are four options  of choice of the coefficients $s(a)$ in \eqref{sign1}:
\be                       \label{opt1}
s(1)=s(2)=\pm 1~~~~~~\mbox{or}~~~~~
s(1)=-s(2)=\pm 1.
\ee
These four options correspond  to  the four 
different branches of the same solution that should be considered all together, 
since for each individual branch functions $X_a,Y_a$ are either discontinuous or not smooth at the cuts. 
However, considering them together 
for all four different values of $s(a)$  gives a smooth Riemann surface
consisting of four sheets 
continuously joining each other through the cuts. Specifically, amplitudes \eqref{sign1} 
are already smooth for $\rho>|\mu_a|$, while for $\rho<|\mu_a|$ 
one should  continue them through the cuts via
(compare with \eqref{anal}) $X_a\to \sign(z-z_a)X_a$ and $Y_a\to \sign(z-z_a)Y_a$ which gives 
\be                         \label{sign2}
X_a=s(a)\,\sign(z-z_a){\cal X}_a,~~~~~Y_a=s(a)\,\sign(\mu_a){\cal Y}_a. 
\ee
These functions are smooth and continuous for $\rho<|\mu_a|$. Using them in \eqref{2w}
gives plots of the solution shown in Fig.\ref{FigU} and Fig.\ref{Figk}. One can see that 
$U(\rho,z)$ is four-valued, hence one needs 
 four different Weyl charts to cover the solution: 
\be
D^{s(1)}_{s(2)}=D^\pm_\pm:\{\rho\geq 0,-\infty<z<\infty\}.
\ee
On each chart the signs of $(X_a,Y_a)$ are chosen according 
to the values of $s(a)$ in 
\eqref{sign1}. 
Each chart has two branch cuts \eqref{cuts1} and 
the complete coordinate atlas covering the  manifold 
is obtained by gluing together the four charts to analytically continue the functions 
through the cuts. 
One glues  the ${\cal I}_1$ cuts on the $D^{+}_{+}$ and  $D^{-}_{+}$
charts by identifying the upper edge of  one cut with the lower edge of the other and vice versa, 
similarly for the ${\cal I}_1$ cuts on the $D^{+}_{-}$ and  $D^{-}_{-}$ charts. 
One also glues ${\cal I}_2$ cuts on the $D^{+}_{+}$ and  $D^{+}_{-}$ charts
and on  the $D^{-}_{+}$ and  $D^{-}_{-}$ charts. 
The resulting atlas covers a manifold ${\cal M}$ with four asymptotic regions connected through the 
wormhole throats. The continuation  through cuts implies that for $\rho<|\mu_a|$
(for example at the symmetry axis $\rho=0$) the field amplitudes are given by \eqref{sign2}.

As a result, expanding the function in \eqref{sign2} at small $\rho$ yields 
\be           \label{XYXY}
X_a+iY_a=s(a)\left(z_a^{\pm}+\frac{\rho^2}{2z_a^\pm}+{\cal O}(\rho^4)\right)
\ee
where $z^{\pm}_a$ is defined in \eqref{ZZZ} and   $s(a)$ is given by \eqref{opt1}. 
Using this 
to compute $U$ in \eqref{2w} gives, if $s(1)=s(2)=\pm 1$,
\be            \label{UU1}
U(\rho=0,z)=\pm\left\{
\sigma_1\arctan\left(\frac{z-z_1}{\m_1}\right)
+\sigma_2\arctan\left(\frac{z-z_2}{\m_2}\right)
\right\},
\ee
and, if $s(1)=-s(2)=\pm 1$,
\be           \label{UU2}
U(\rho=0,z)=\pm\left\{
\sigma_1\arctan\left(\frac{z-z_1}{\m_1}\right)
-\sigma_2\arctan\left(\frac{z-z_2}{\m_2}\right)
\right\},
\ee
which are the values of $U$ on the four symmetry axes. This  can be compared with 
Eqs.\eqref{v1},\eqref{v2} in the single ring case, in which case there are only two symmetry axes.

Let us now similarly compute the value of $k$ at the symmetry axes. 
The amplitudes $k_a$ in \eqref{2w2} are insensitive to signs of $X_a$, $Y_a$
hence for given $(\rho,z)$ their values are the same on all  Weyl charts. 
The $k_{12}$ amplitude in \eqref{2w2} depends on products 
$(X_1+iY_1)(X_2+iY_2)$   and $(X_1+iY_1)(X_2-iY_2)$ for which there are two sign options. 
If one chooses in \eqref{XYXY} $s(1)=s(2)$  then one will have for small $\rho$
$$
(X_1+iY_1)(X_2+iY_2)=z_1^{+}z_2^{+}+{\cal O}(\rho^2), ~~~~
(X_1+iY_1)(X_2-iY_2)=z_1^{+}z_2^{-}+{\cal O}(\rho^2), 
$$
inserting which to \eqref{2w1},\eqref{2w2} gives 
\be
k_1(0,z)=k_2(0,z)=k_{12}(0,z)=0.
\ee
The other possibility is to chose in \eqref{XYXY} $s(1)=-s(2)$, then one has at small $\rho$
\bea
(X_1+iY_1)(X_2+iY_2)=-\left({z}_1^+ +\frac{\rho^2}{2 z_1^{+}  }\right)\left(z_2^{+}+\frac{\rho^2}{2 z_2^{+}}\right)
+{\cal O}(\rho^4),
\nonumber \\
(X_1+iY_1)(X_2-iY_2)=-\left({z}_1^+ +\frac{\rho^2}{2 z_1^{+}  }\right)\left(z_2^{-}+\frac{\rho^2}{2 z_2^{-}}\right)
+{\cal O}(\rho^4).
\eea
Inserting this to \eqref{2w1},\eqref{2w2} gives 
\be
k_1(0,z)=k_2(0,z)=0,
\ee
but $k_{12}$ assumes a constant non-zero value  
\be                  \label{kkk}
k_{12}(0,z)=\ln\left(
\frac{(z_1-z_2)^2+(\m_1-\m_2)^2}{(z_1-z_2)^2+(\m_1+\m_2)^2}
\right). 
\ee
The conclusion is that $k(\rho,z)$ vanishes at the two symmetry axes with $s(1)=s(2)$ 
where $U$ is given by \eqref{UU1}, hence these two  axes are regular. 
However, $k$ assumes the constant non-zero value \eqref{kkk} 
everywhere at the other two axes, where $s(1)=-s(2)$ and $U$ is given by \eqref{UU1}.
Hence these two axes are singular and contain infinite struts (see Fig.\ref{Figk}). 
Therefore, two of the four symmetry axes of the solution 
are  regular while the other two are everywhere singular. 

The two-wormhole system was  first discussed  in \cite{Clement:1983tu}, 
where  the solution for $U(\rho,z)$  was obtained (see also \cite{Egorov:2016rfr}). 
It took next   more than 30 years to obtain also the explicit 
solution for $k(\rho,z)$  and not just the integral representation \eqref{kkkk} \cite{Clement:2015lul}. 
The formula \eqref{kkk} was also obtained in \cite{Clement:2015lul}, however, 
it was concluded there that $k(0,z)$ is 
discontinuous  and assumes the value \eqref{kkk} either inside 
the interval $[z_1,z_2]$ while vanishing outside, or the other way round. 
As one can see in Fig.\ref{Figk} (left panel)
precisely this type of behaviour is shown {\it separately}  by the two (blue and yellow online) 
solution branches.
However, $k$ is clearly continuous on the combined surface made of both branches. 
Moreover, choosing regular near the axis amplitudes  $X_a,Y_a$ \eqref{sign2} 
renders $k(0,z)$  constant for each separate branch,
as one can see in the right panel of  Fig.\ref{Figk}.  In any case, since $U$ is everywhere finite and smooth, 
the field equations imply that 
the derivative $\partial_z k\sim \rho$ should vanish at the axis hence $k(0,z)$ cannot jump. 

\subsection{Locally flat wormholes}

Summarising the above discussion, the oblate vacuum solutions obtained from the two-rod metrics 
describe a pair of wormholes. An analysis similar to that in Section \eqref{rwh} above shows that their curvature 
exhibits the conical and power-law singularities at the two circles $(\rho,z)=(\mu_a,z_a)$, which can be 
viewed as the positions of singular matter sources -- two negative tension rings. 
 In addition, the solutions  show 
 a conical singularity everywhere at the two of their four symmetry axes where the function $k$ assumes a constant
non-zero value. Therefore, these solutions are singular even at infinity.

\begin{figure}[th]
\hbox to \linewidth{ \hss

	\resizebox{10cm}{10cm}{\includegraphics{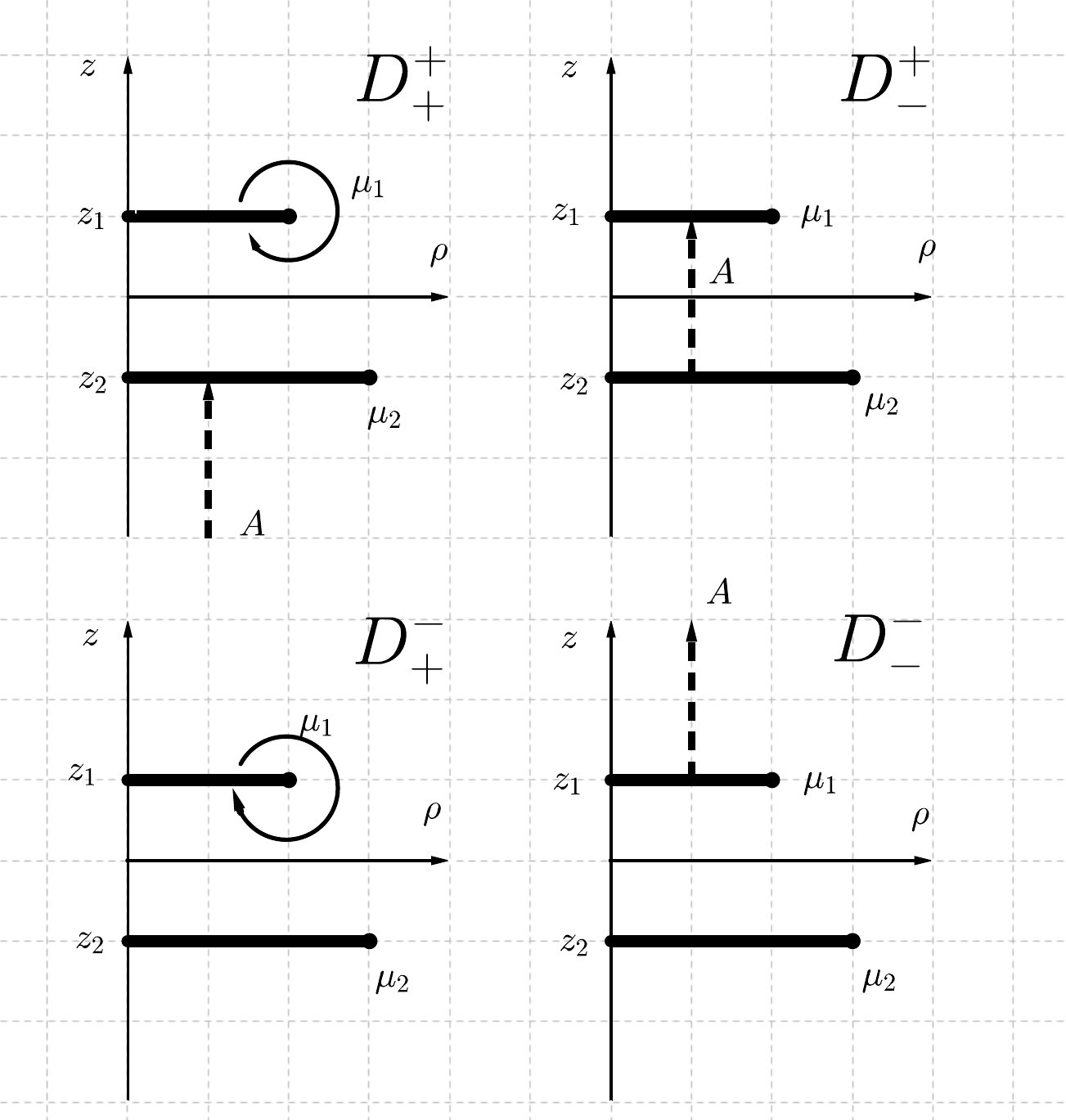}}
	
	
\hss}
\caption{The two-ring wormhole is covered by four Weyl charts, each having two branch cuts. 
The upper cuts on $D^{+}_\pm$ and $D^{-}_\pm$ are  glued to  each other such that the upper
edge of the one is identified with the lower edge of the other and vice-versa; similarly 
for the lower cuts on $D^{\pm}_{+}$ and $D^{\pm}_{-}$. As a result, the geodesic $A$ crosses
several charts.  
}
\label{FFF}

\end{figure}

There is, however, a notable exception  obtained by taking in \eqref{2w} the $\sigma_1\to 0$ and 
$\sigma_2\to 0$ limit, which gives $U=k=0$ everywhere, hence the metric becomes locally flat. 
The conical singularity along the $z$-axes 
as well as the power-law curvature singularity at the rings then disappear. However,  the topology remains 
non-trivial as one still needs four Weyl charts to cover the manifold, hence the 
conical singularities at the points $(\mu_a,z_a)$ of the $(\rho,z)$ plane remain 
(we now assume without loss of generality that $\mu_a>0$ since in the opposite case one should simply replace in all
formulas below $\mu_a$ by $|\mu_a|$). 
For example, a contour around the point $(\mu_1,z_1)$ in the $D^{+}_{+}$ patch shown in Fig.\ref{FFF} does not 
close after the first revolution 
since when arriving at the lower edge of the cut it passes  to the $D^{-}_{+}$ patch and only after a second revolution 
returns back to the $D^{+}_{+}$ patch to close. 
Therefore, the total angle increment  is $4\pi$ hence there is an angle deficit of 
$-2\pi$ corresponding to a cosmic string of negative tension $T=-c^4/(4G)$.
This string stretches in the $\varphi$-direction, hence this is a ring of radius $\mu_1$
placed at $z=z_1$. 
Similar argument shows that there is also a  ring of the same tension and 
of radius $\mu_2$ placed at $z=z_2$. 
Therefore, the solution is sourced by a pair of negative tension rings 
whose positions and radii can be arbitrary. 

Since the metric is locally flat, the geodesics are simply straight lines. Those 
initially parallel to the $z$-axis will always remain parallel to it, which helps to understand once again  why 
there are four symmetry axes. Consider a geodesic which is close enough to the axis to thread both rings,
as for example the geodesic $A$ in Fig.\ref{FFF} which stars at the $D^{+}_{+}$ chart at $z=-\infty$. 
When it threads the ring at $z=z_2$ it passes to the $D^{+}_{-}$ chart and continues there, until it threads 
the second ring at $z=z_1$ and passes to the $D^{-}_{-}$ chart where it stays till reaching $z=+\infty$. 
Therefore, this geodesic follows the pattern 
\be
D^{+}_{+}\to D^{+}_{-} \to D^{-}_{-}\,
\ee
which determines one of the four $z$-axes of the solution. Similarly, a geodesic which starts on $D^{-}_{-}$ 
and follows the $z$-axis proceeds as 
\be
D^{-}_{-}\to D^{-}_{+} \to D^{+}_{+}\,
\ee
which determines the second axis. The other two axes are similarly defined by 
\be
D^{+}_{-}\to D^{+}_{+} \to D^{-}_{+}\,,~~~~~~~
D^{-}_{+}\to D^{-}_{-} \to D^{+}_{-}\,. 
\ee

The above arguments can be straightforwardly generalized to the case of $N$ rings 
of arbitrary radii $\mu_a$ and positions $z_a$ and of the same tension $T=-c^4/(4G)$. 
The metric expressed in Weyl coordinates is manifestly flat but the topology is non-trivial and one needs
$2^N$ Weyl charts 
\be
D_{s(1)s(2)\ldots s(N)}:\{\rho\geq 0,-\infty<z<\infty\}
\ee
to cover the whole of the manifold. Here $s(a)=\pm 1$ and each chart has $N$ cuts 
\be
{\cal I}_a=\left\{\rho\in[0,\mu_a],z=z_a\right\},~~~~a=1,\ldots N.
\ee
The atlas is obtained by gluing together the $m$-th cuts on all pairs of charts 
with coinciding indices $s(a\neq m)$. 
For example, the upper edge of the first cut on $D_{+s(2)\ldots s(N)}$ should be identified 
with the lower edge of the first cut on $D_{-s(2)\ldots s(N)}$ and vise versa for all possible values of 
$s(2),\ldots s(N)$. This gives an exact solution of Einstein equations sourced by $N$ singular rings
whose  geometry is everywhere flat except at the ring positions.  

\subsection{Solutions with scalar field}

Any of the discussed above vacuum Weyl metrics described by 
\be              \label{ss0}
(U,k,\Phi=0)
\ee
can be promoted to solutions with $\Phi\neq 0$.
Applying symmetries \eqref{sym1} and \eqref{sym2} gives
solutions with the conventional scalar, 
\be                  \label{ss1}
U_s=\frac{1}{s }\,U,~~~~~k_s=k,~~~~\phi_s=\frac{\sqrt{s^2-1}}{s}\,U, ~~~~s\geq 1,
\ee
boosted solutions with phantom field, 
 \be                 \label{ss2}
U_s=\frac{1}{s}\,U,~~~~~k_s=k,~~~~\psi_s=\frac{\sqrt{1-s^2}}{s}\, U, ~~~~s\leq 1,
\ee
and their swapped version 
 \be                 \label{ss3}
U_s=\frac{1}{s}\,U,~~~~~k_s=-k,~~~~\psi_s=\frac{\sqrt{1+s^2}}{s}\,U, ~~~~s\in(-\infty,\infty).
\ee
Since the original solution can exist in prolate and oblate versions, this gives six one-parameter 
families of solutions with scalar. In addition, 
twice as many solutions can be obtained by  acting on \eqref{ss1}--\eqref{ss3}
with the tachyon symmetry \eqref{tachyon}.

For example, taking the two-wormhole vacuum metric \eqref{2w}, 
applying the swap symmetry \eqref{ss3}, and then taking the limit $s\to\infty$ 
gives the two-wormhole counterpart of the ultrastatic solution of Bronnikov and Ellis.  
This solution is no longer spherically symmetric and  it contains infinite  struts 
along two symmetry axes. 

More generally, all solutions with $\Phi\neq 0$ obtained from the two-rod metrics are singular 
since already their prolate and oblate vacuum versions have struts. 
The only solutions without struts 
are the vacuum rings with $U=k=0$, but symmetries \eqref{ss1}--\eqref{ss3} act trivially in this case
and give $\Phi=0$.

\section{Solutions from point masses -- Appell wormhole\label{apl}}

Let us finally  discuss solutions obtained from the Chazy-Curzon metric \eqref{Curz},\eqref{Curz1}. 
This metric  does not admit the ``prolate'' and ``oblate'' generalizations since the scale symmetry 
$(U,k)\to(\lambda U,\lambda^2 k)$ acts trivially only redefining 
 the parameter $m$ in \eqref{Curz}. Hence, there is only one vacuum solution, 
\be                        \label{Curz0}
U= - \frac{m}{R}\,, \qquad
 k =-\frac{m^2\rho ^2}{2 R^4 } \,,
\ee
with $R=\sqrt{\rho^2+z^2}$, inserting which to \eqref{ss1}--\eqref{ss3} gives 
solutions with scalar field. 

More possibilities exist for the two mass  solution \eqref{Curz1} because it contains three free parameters, 
$m_\pm$ and $m$. Applying $(U,k)\to(\lambda U,\lambda^2 k)$ again results in a 
trivial rescaling of $m_\pm$,  but this time there is a non-trivial possibility to complexify  via 
\be
m\to i\mu,~~~~~~m_\pm\to -\frac{M}{2}\,e^{\pm i\eta}\,, 
\ee
with constant and real $\mu,M,\eta$. 
This gives 
\be
R_{\pm}=\sqrt{\rho^2+(z\pm m)^2}
&\to& \sqrt{\rho^2+(z\pm i\mu)^2}=X\pm iY = x\pm i\m\cos\vartheta \equiv {\cal R}e^{\pm i{\cal S}}~~ 
\ee
with ${\cal R}=\sqrt{X^2+Y^2}$ 
and
$
\tan({\cal S})={Y}/{X}
$
where $X,Y$ in terms of $\rho,z$ are given by 
\eqref{plus}. 
Applying this to  \eqref{Curz1} gives the real-valued solution, 
\be                      \label{Curz2}
U=\frac{M}{\cal R}\,\cos({\cal S}-\eta),~~~~k=-\frac{M^2\rho^2}{4{\cal R}^4}\,\cos(4{\cal S}-2\eta)
-\frac{M^2}{8\mu^2}\left(\frac{\rho^2+z^2+\m^2}{{\cal R}^2}-1\right).
\ee
This is regular at the symmetry axis,
$
k(0,z)=0,
$
and  describes the so called Appell ring whose Newtonian 
potential $U$ \cite{Appell}  is produced by a disk of negative mass density 
whose rim has positive mass density 
\cite{App,Letelier:1997dz,Semerak:2016dqd}. 
Since the phase ${\cal S}$ is double-valued around the branching point $(\rho,z)=(\mu,0)$,
the solution has the same double-sheeted topology as the considered above wormholes. Hence 
this is again a wormhole. 
However, since ${\cal R}$ vanishes at the ring, the Newtonian potential diverges \cite{Semerak:2016dqd}. 

Taking as the pair $(U,k)$ either the original  Chazy-Curzon solution \eqref{Curz1}
or the Appell ring \eqref{Curz2}  and injecting to \eqref{ss1}--\eqref{ss3}
gives  solutions with scalar field. 
In particular, applying to  \eqref{Curz2}  the boost  \eqref{ss2} 
and then taking the infinite boost limit $s\to 0$ while keeping constant the ratio 
${M}/{s}\equiv A$  
yields $k=0$ and 
\be
U=\psi=A\, \frac{\cos({\cal S}-\eta)}{\cal R}=A\cos\eta\,\frac{x}{x^2+\mu^2\cos^2\vartheta}
+A\mu\sin\eta \,\frac{\cos\vartheta}{x^2+\mu^2\cos^2\vartheta}. 
\ee
For $\eta=\pi/2$ this reduces to the ring solution of \cite{Miranda:2013gqa,Matos:2012gj} 
discussed around Eq.\eqref{Uap}.

\section{Conclusions}

To recapitulate, we applied above the duality rotations to the 
vacuum Weyl metrics to produce new solutions with scalar field. We were mainly  interested in 
the wormhole solutions with  several asymptotically flat  regions. 
Such solutions are best known in systems with exotic matter, but it is little known that 
they exist also in vacuum GR, being sourced by  thin negative tension rings. 
The one-ring solutions show   a conical singularity and a power-law curvature 
singularity at the ring, but the Newtonian potential is finite everywhere. 
In this respect, the ring wormholes differ  from other known solutions with 
rings/strings which show infinite red or blue shifts  or 
pathologies like closed timelike curves. 
These are, for example, 
the Kerr black hole containing inside the horizon a ring singularity which 
can also be viewed as a wormhole \cite{Carter:1968rr}, or the   Appell ring considered above
(see  \cite{Semerak:2016dqd} for other singular rings), 
or the NUT wormholes
containing  inside a singular Misner string  \cite{Clement:2015aka,Ayon-Beato:2015eca}. 

The vacuum ring wormholes  could be viewed as ``primary"  while 
solutions with scalar field  are obtainable from them via duality rotations. 
This gives rings ``dressed with scalar", conventional or phantom. In particular, 
the spherically symmetric BE wormhole can be obtained by dressing the ring with the phantom field. 
As we have seen, one can construct large families of  such ``dressed up" solutions.

The most interesting property of our ring wormholes  is that, unlike other known rings, 
they admit a remarkable limit where their geometry becomes locally flat -- when 
the tension of all rings  attains the  value $T=-c^4/(4G)$. 
The topology then remains non-trivial and shows several interconnected asymptotic regions. 
Solutions then become regular everywhere apart from the rings where they show a mild 
conical singularity, which could presumably be smoothened by replacing the thin rings by 
regular toroidal sources of finite thickness. Such rings literally cut holes in flat space.

In some parts of our discussion  we
summarised and generalised little known facts  scattered in the literature, 
whereas other parts  are original, as for example the description 
of  the ``scalar-dressed"  solutions obtained via duality transformations and of the 
ring and multi-ring solutions with locally flat geometry. The main message we were trying
to convey is that traversable wormholes could be less exotic objects than is usually thought, 
since they exist already in vacuum GR and 
not necessarily only in systems with exotic matter.

\section*{Acknowledgements} 
We thank G\'erard Cl\'emant for discussions. 
G.W.G. thanks the LMPT for hospitality and acknowledges the support of  ``Le Studium" -- Institute
for Advanced Studies of the Loire Valley. 
M.S.V. was partly supported by the Russian Government Program of Competitive Growth 
of the Kazan Federal University.

\appendix
\setcounter{section}{0}
\setcounter{equation}{0}
\setcounter{subsection}{0}
\section{Isometric embeddings of the BE wormhole  \label{A}}

\renewcommand{\theequation}{\Alph{section}.\arabic{equation}} 
Introducing 
$
(X_1,X_2,X_3)=\sqrt{x^2+\mu^2}\,(\sin\vartheta \cos\varphi,\sin\vartheta\sin\varphi,\cos\vartheta)
$
and also $Z$, 
\be                            \label{dZ}
dZ=\frac{\mu}{\sqrt{x^2+\mu^2} }\,dx~~~~~\Rightarrow~~~~~x=\sinh\left(\frac{Z}{\mu}\right),
\ee
the ultrastatic BE solution \eqref{UW} can be represented as the  geometry induced on a 
hypersurface  
in five dimensional Minkowski space,  
\be
ds^2=-dt^2+dx^2+(x^2+\mu^2)d\Omega^2=-dt^2+dX_1^2+dX_2^2+dX_3^2+dZ^2\,,
\ee
where 
\be
X_1^2+X_2^2+X_3^2=\mu^2 \cosh^2\left(\frac{Z}{\mu}\right). 
\ee
\begin{figure}[th]
\hbox to \linewidth{ \hss

	\resizebox{8cm}{5cm}{\includegraphics{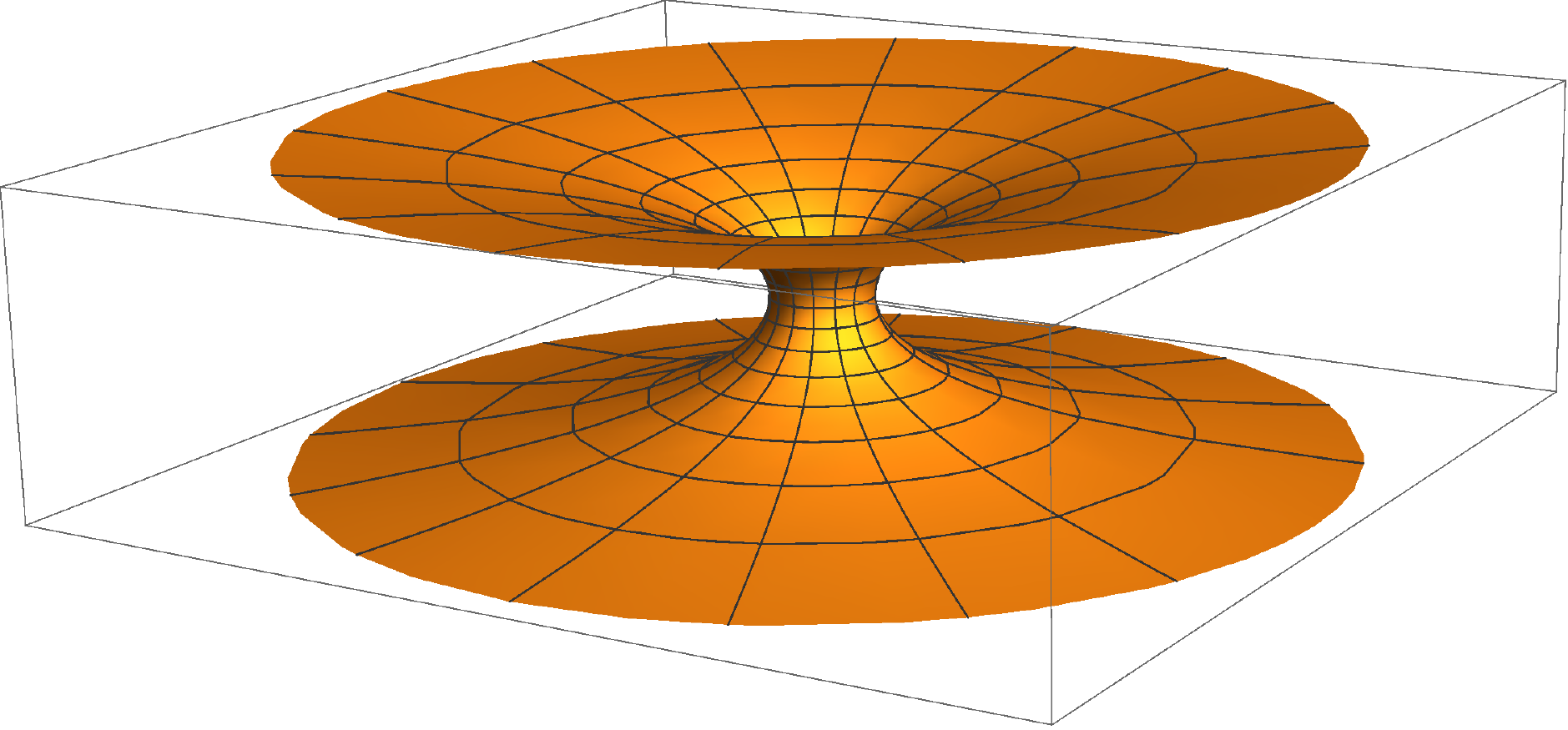}}
\hspace{1mm}

\hss}
\caption{Embedding 
of the equatorial section of the BE wormhole to the 3-dimensional Euclidean 
space spanned by $X_1,X_2,Z$ defined by \eqref{BE1},\eqref{BE2}.}
\label{FigBE}
\end{figure}
It follows that the spatial 2-geometry of the equatorial plane $X_3=0$  is a catenoid (see Fig.\ref{FigBE}) -- 
a surface of revolution in three-dimensional Euclidean space with the line element 
\be                \label{BE1}
dX_1^2+dX_2^2+dZ^2=
dr^2+r^2d\varphi^2+dZ^2
\ee
whose meridional curve is the catenary, 
\be               \label{BE2}
r=\mu \cosh\left(\frac{Z}{\mu}\right). 
\ee
One can also construct the embedding of the non-ultrastatic BE geometry \eqref{W},
\bea                                  \label{WA}
ds^2= -A^2(x) dt^2+\frac{dx^2}{B^2(x)}+r^2(x)(d\vartheta^2+\sin^2\vartheta d\varphi^2)
\eea
with 
\be                  \label{rrr}
A(x)=B(x)=e^{\Psi/s},~~~~~~~r(x)=\frac{\sqrt{x^2+\mu^2}}{B(x)},~~~~~~\Psi=\arctan\left(
\frac{x}{\mu}
\right),
\ee
following the procedure of \cite{Fronsdal:1959zza,Ferraris}.  
This geometry can be embedded into the 7-dimensional Minkowski 
space with the metric 
\be                     \label{amb} 
ds^2=-dX_0^2+dX_1^2+dX_2^2+dX_3^2+dX_4^2+dX_5^2-dX_6^2
\ee
by the following explicit formulas for  $X_k=X_k(t,x,\vartheta,\varphi)$:
\be
X_0&=&A(x)\sinh (t),~~~~~~X_1= A(x)\cosh (t),~~~~~~~\nonumber \\
X_2&=&r(x)\sin\vartheta\cos\varphi,~~~X_3=r(x)\sin\vartheta\sin\varphi,~~~
X_4=r(x)\cos\vartheta, ~~~~\nonumber \\
X_5&=&\int\frac{dx}{B(x)},~~~~~~X_6=\int\sqrt{A^{\prime 2}(x)+r^{\prime 2}(x)  }\,dx\,,
\ee
where the prime denotes the derivative with respect to $x$. 

\appendix
\setcounter{section}{1}
\setcounter{equation}{0}
\setcounter{subsection}{0}
\section{Flamm embedding versus  Einstein-Rosen bridge  \label{A0}}

\renewcommand{\theequation}{\Alph{section}.\arabic{equation}} 
It is instructive  to discuss the relation between the work of Flamm  \cite{Flamm} of 1916 
and  analysis of Einstein and Rosen \cite{Einstein:1935tc} of 1935 presenting the first ever example of wormholes. 

Flamm (see \cite{Flamm2015},\cite{FlammGibbons} for the English translation)
was the first to consider isometric embeddings 
of the Schwarzschild solution. Specifically, the spatial part of the 
vacuum Schwarzschild metric in the $r>2M$ region can be represented as 
\be
dl^2=\frac{dr^2}{1-2M/r}+r^2d\Omega^2=dr^2+r^2d\Omega^2+dZ^2
\ee
where 
\be                       \label{A2}
dZ^2=\frac{dr^2}{{r/(2M)-1}}~~~~~\Rightarrow~~~~~
r=r(Z)\equiv 2M+\frac{Z^2}{8M}.
\ee
\begin{figure}[th]
\hbox to \linewidth{ \hss

	\resizebox{7cm}{6cm}{\includegraphics{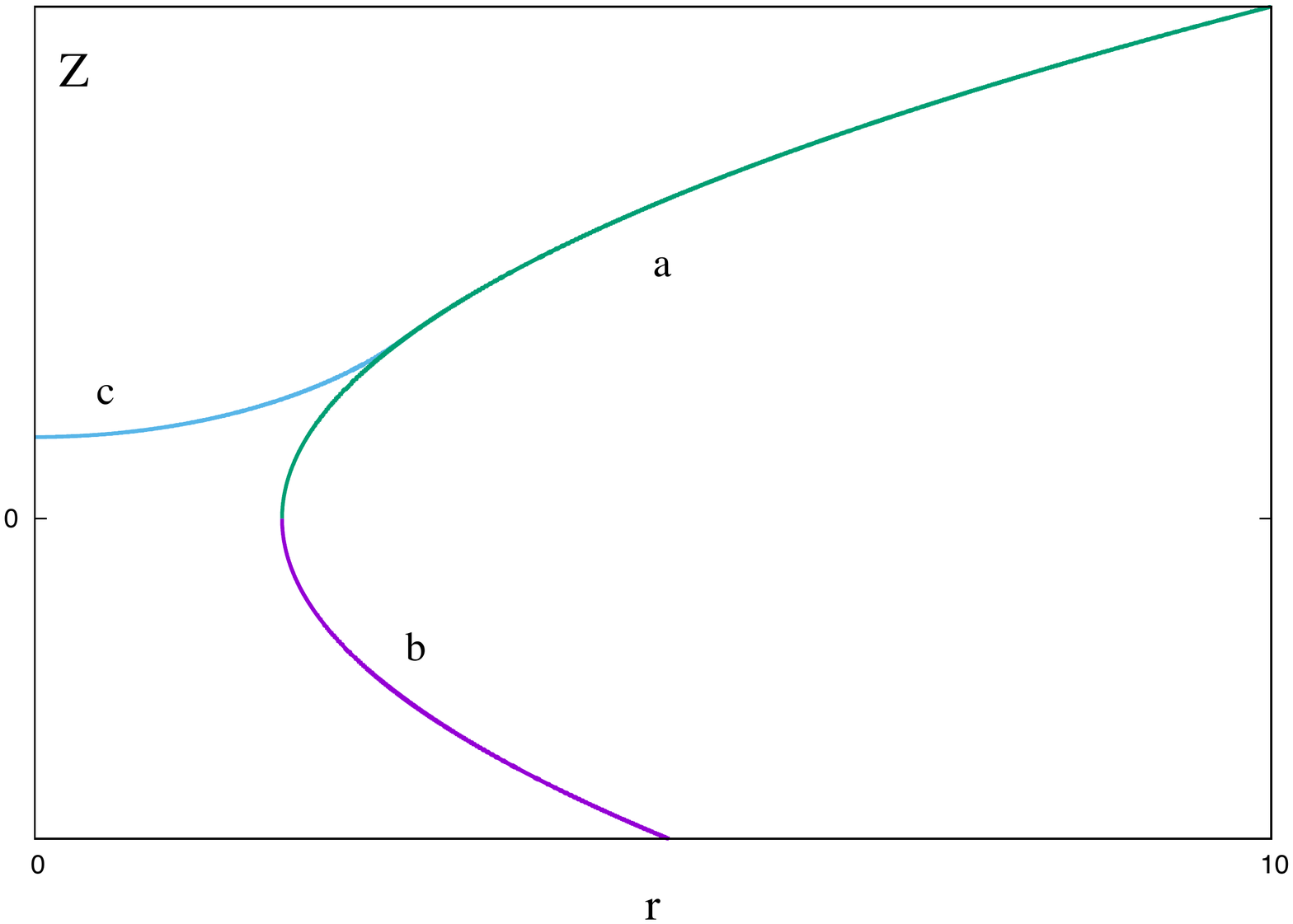}}
\hspace{3 mm}
	\resizebox{9cm}{6cm}{\includegraphics{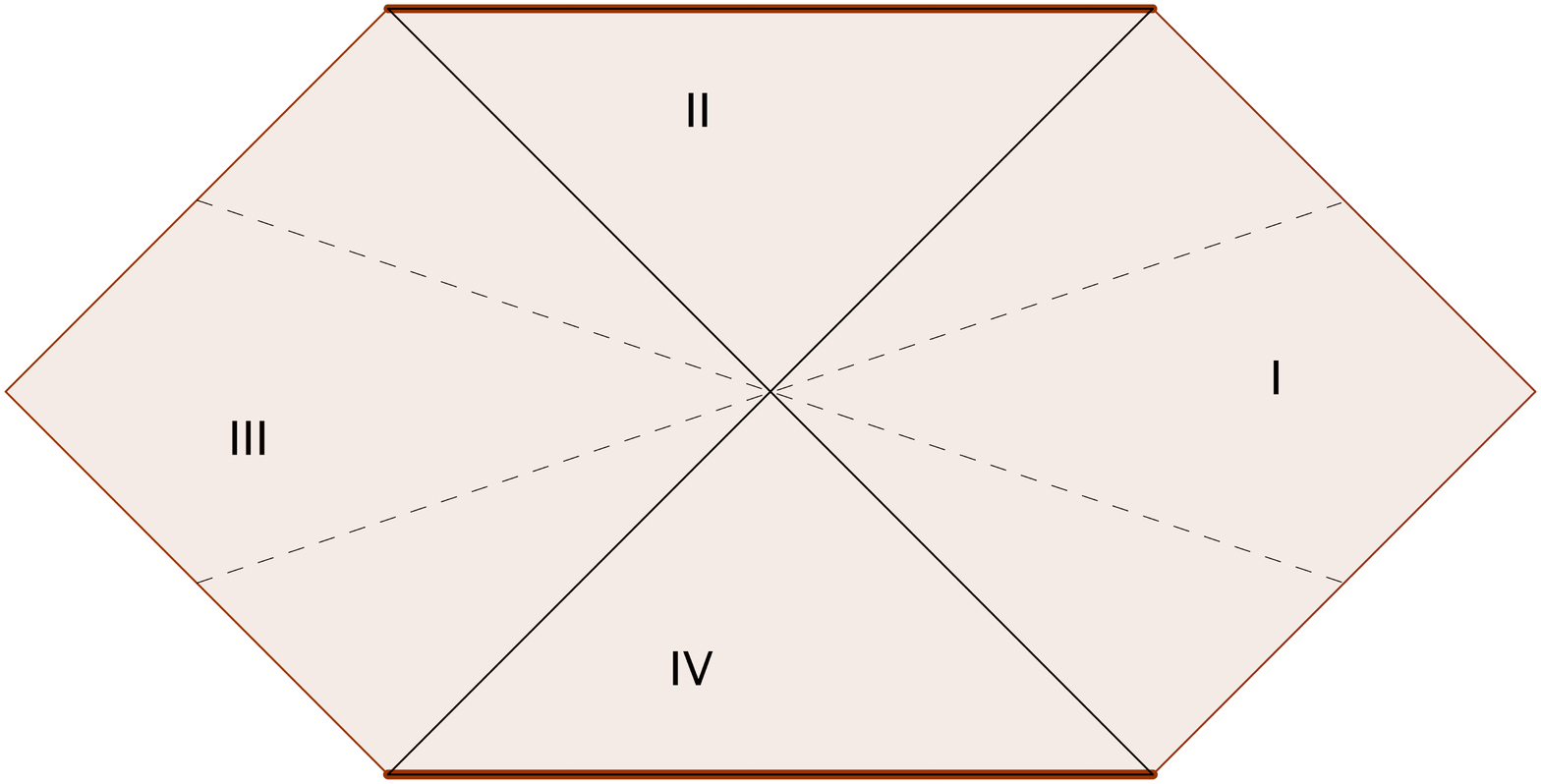}}

\hss}
\caption{Left: the Flamm embedding -- the surface obtained by rotating 
the curve $c$
(interior region) and the smoothly joining it 
part of the curve $a$ (exterior region) 
around the $Z$-axis. 
Right: the Einstein-Rosen  
coordinates $t,Z$ covering the $I+III$ regions of the Kruskal manifold. 
$Z$ varies from 
$-\infty$ to $+\infty$ along the dashed lines of constant $t$. The bridge is at the central $Z=0$ point 
joining the two regions. }
\label{Fl}
\end{figure}

Therefore, the geometry is the same as that induced on the paraboloid 
of revolution  obtained by rotating the parabola 
$r=r(Z)$   around the Z-axis of the four-dimensional Euclidean space. 
The parabola consists of two parts denoted by $a$ and $b$ in Fig.\ref{Fl}
which belong, respectively, to the  $Z>0$ and $Z<0$ regions. 
Since the $a$-part is enough 
to cover the $r>2M$ region, Flamm retains only this curve. It smoothly  matches 
at a point $r=r_0>2M$  the curve $c$ 
describing  the embedding of the 
{\it interior} solution  with  a matter source in  the $r<r_0$ region (see Fig.\ref{Fl}). 
Therefore,  the surface obtained by rotating the curves $a+c$ in Fig.\ref{Fl} around 
the Z-axis has the same 3-geometry as that for the combined exterior+interior Schwarzschild solution. 
The curve $b$ in Fig.\ref{Fl}  can  be disregarded in this case. 

Einstein and Rosen (ER) \cite{Einstein:1935tc} did not refer to Flamm  but their construction  can 
be easily understood using  Flamm's analysis. Specifically, they retain the curve $b$ 
but disregard   the $r<2M$ region by setting 
in the line element $r=r(Z)$ with $Z\in(-\infty,+\infty)$ becoming 
the radial coordinate (they actually  use the variable $u=Z/\sqrt{8M}$). 
Rotating 
the full parabola $a+b$ around  the $Z$-axis then gives a surface 
similar to the wormhole shown  in Fig.\ref{FigBE}. 
Using the modern formulation in terms of the 
Kruskal   extension, the  ER coordinates 
cover  the $I$ and $III$  regions  of the conformal
diagram shown in Fig.\ref{Fl}. 
These two exterior parts are connected by the bridge (wormhole thorat) -- the surface of 
minimal radius $r(0)=2M$ corresponding to the central point of the diagram (the Boyer axis). 
The bridge cannot be traversed by 
ordinary matter  because  intervals between points in region $I$ and those 
in region $III$ are spacelike. 

One should stress that, although the 3-metric is regular, the full 4-metric 
\be
ds^2=-\frac{Z^2}{16M^2+Z^2}\,dt^2+\left(1+\frac{Z^2}{16M^2}\right)dZ^2+
4M^2\left(1+\frac{Z^2}{16M^2}\right)^2 d\Omega^2
\ee
degenerates at $Z=0$ hence the ER coordinates are not global. 
At the same time, replacing $Z^2\to Z^2+\alpha^2$ with a small constant $\alpha$ 
in the numerator of $g_{00}$ would produce a globally static solution 
with a thin shell of matter around $Z=0$ resembling  a thin shell wormhole
in modern language.
This is similar to some ideas expressed in the ER paper.

Contrary to what one often sees in the literature, the Flamm-ER 3-surface 
{\it does not} become flat for $r\to\infty$ since its meridional parabola does not have a linear asymptote.






\providecommand{\href}[2]{#2}\begingroup\raggedright\endgroup

\end{document}